\documentclass{article}

\usepackage{arxiv}
\usepackage[utf8]{inputenc}
\usepackage{amsfonts,amssymb,amsmath}
\usepackage{hyperref}
\usepackage[numbers]{natbib}
\newtheorem{theorem}{Theorem}
\newtheorem{definition}{Definition}

\usepackage{tikz}
\usetikzlibrary{shapes, fit, positioning, calc}
\tikzset{
	double color fill/.code 2 args={
		\pgfdeclareverticalshading[%
			tikz@axis@top,tikz@axis@middle,tikz@axis@bottom%
		]{diagonalfill}{100bp}{%
			color(0bp)=(tikz@axis@bottom);
			color(50bp)=(tikz@axis@bottom);
			color(50bp)=(tikz@axis@middle);
			color(50bp)=(tikz@axis@top);
			color(100bp)=(tikz@axis@top)
		}
		\tikzset{shade, left color=#1, right color=#2, shading=diagonalfill}
	}
}

\usepackage{xcolor}

\newcommand{\eps}{\ensuremath{\epsilon}}
\newcommand{\dlt}{\ensuremath{\delta}}
\newcommand{\sgm}{\ensuremath{\sigma}}

\newcommand{\cali}[1]{\ensuremath{\mathcal{#1}}} 
\newcommand{\var}[1]{\ensuremath{\mathtt{#1}}}

\usepackage{todonotes}
\usepackage{csquotes}
\usepackage{comment}
\usepackage{multirow}
\usepackage{subcaption}
\usepackage{adjustbox}
\usepackage{tabularx}

\title{On the privacy-utility trade-off in differentially private hierarchical text classification}

\author{
  Dominik Wunderlich, Daniel Bernau, Francesco Aldà \\
  SAP\\
  Karlsruhe, Germany \\
  \texttt{firstname.lastname@sap.com} \\
   \And
 Javier Parra-Arnau \\
  Karlsruhe Institute of Technology\\
  Karlsruhe, Germany\\
  \texttt{javier.parra-arnau@kit.edu} \\
   \And
 Thorsten Strufe \\
  Karlsruhe Institute of Technology\\
  Karlsruhe, Germany\\
  \texttt{strufe@kit.edu} \\
}

\begin{document}

\maketitle

\begin{abstract}
{Hierarchical text classification consists in classifying text documents into a hierarchy of classes and sub-classes. Although artificial neural networks have proved useful to perform this task, unfortunately they can leak training data information to adversaries due to training data memorization. Using differential privacy during model training can mitigate leakage attacks against trained models, enabling the models to be shared safely at the cost of reduced model accuracy.
This work investigates the privacy-utility trade-off in hierarchical text classification with differential privacy guarantees, and identifies neural network architectures that offer superior trade-offs.
To this end, we use a white-box membership inference attack to empirically assess the information leakage of three widely used neural network architectures. We show that large  differential privacy parameters already suffice to completely mitigate membership inference attacks, thus resulting only in a moderate decrease in model utility. More specifically, for large datasets with long texts we observed Transformer-based models to achieve an overall favorable privacy-utility trade-off, while for smaller datasets with shorter texts convolutional neural networks are preferable.}
\end{abstract}

\keywords{Anonymization, Text Classification}
\section{Introduction}
\label{sec:intro}
Organizing large corpora of unstructured data such as text documents, news articles, emails, and support tickets in an automated manner is a considerable challenge due to the inherent ambiguity of natural languages~\cite{HFB19JPC}. However, the automated classification of unstructured data overcomes manual data labelling activities and thus is a key capability for organizing data at scale~\cite{TaylorWIRED13}. Due to the wide range of applications, Hierarchical Text Classification (HTC) has received particular interest by the Natural Language Processing (NLP) community in recent years~\cite{mao-etal-2019-hierarchical, qu2012evaluation, agrawal2013multi, peng2016deepmesh}. HTC leverages machine learning to automatically organize documents into taxonomies, predicting multiple labels in a predefined label hierarchy.

After data owners have trained HTC models on their data the models may be shared with data analysts such as contractors, customers, or even the general public. 
However, sharing a model may leak information about the training data~\cite{shokriMembershipInferenceAttacks2017, nasrComprehensivePrivacyAnalysis2018, zhangUnderstandingDeepLearning2017, carliniSecretSharerEvaluating2019}. Perturbation with differential privacy (DP)\footnote{For conciseness,
throughout this work we use the acronym DP to refer to both ``differential privacy'' and its adjective form ``differentially private''.} limits information leakage by anonymizing the training data or model training function~\cite{dworkDifferentialPrivacy2006,abadiDeepLearningDifferential2016,Hayes2019}. DP introduces an inherent trade-off between privacy and utility, which means that a stronger privacy guarantee implies a decrease in informative value. Balancing this trade-off is especially hard when training Artificial Neural Networks (ANNs). On the one hand, ANN utility can only be assessed empirically after training and even small perturbation can have a high impact on utility~\cite{BPS19}. 
On the other hand, DP anonymization parameter $\epsilon$ formulates a theoretic upper bound on information leakage that does not reflect the empirical information leakage for a concrete dataset. For ANNs empirical information leakage can be assessed with Membership Inference (MI) attacks~\cite{shokriMembershipInferenceAttacks2017}, which aim at identifying single instances of the training data by sole access to the trained model~\cite{nasrComprehensivePrivacyAnalysis2018, rahmanMembershipInferenceAttack2018}. A rather large gap has been observed between the high theoretical bound on information leakage under MI attacks that can be derived from DP guarantees and the empirical information leakage posed by MI attacks~\cite{jayaramanEvaluatingDifferentiallyPrivate2019}. Consequently, choosing the anonymization strength via privacy parameter $\varepsilon$ remains a challenging problem since a data owner can either choose to consider the theoretical or empirical information leakage. 

Our work compares the empirical privacy-utility trade-off of multiple HTC models by quantifying privacy under MI. We hypothesize that there are preferable ANN architectures w.r.t.~the privacy-accuracy trade-off when applying DP to HTC. Unlike previous studies on the privacy-accuracy trade-off for numerical or image data~\cite{rahmanMembershipInferenceAttack2018,bernauAssessingDifferentiallyPrivate2020, jayaramanEvaluatingDifferentiallyPrivate2019}, our work focuses on textual data which requires different ANN architectures, such as Transformers~\cite{vaswaniAttentionAllYou2017}. The main contributions of this paper are:
\begin{itemize}
    \item Empirically quantifying and comparing the privacy-utility trade-off for three widely used HTC ANN architectures on three reference datasets. In particular, we consider Bag of Words (BoW), Convolutional Neural Networks (CNNs) and Transformer-based architectures.
    \item Connecting DP privacy guarantees to MI attack performance for HTC ANNs. In contrast to the adversary considered by DP, the MI adversary represents an ML specific threat model.
    \item Recommending HTC model architectures and privacy parameters for the practitioner based on the privacy-utility trade-off under DP and MI.
\end{itemize}

This paper is structured as follows. Section~\ref{sec:preliminaries} recalls key aspects of DP, MI attacks and HTC.
Section~\ref{sec:methodology} formulates our approach for modelling the privacy-utility trade-off in HTC. Section~\ref{sec:data} introduces reference datasets and Section~\ref{sec:evaluation:setup} the experiment setup. Experiment results results are presented in Section~\ref{sec:evaluation}  and subsequently discussed in Section~\ref{sec:discussion}. Section~\ref{sec:related_work} introduces related work. Conclusions are drawn in Section~\ref{sec:conclusion}.
\section{Preliminaries}
In the following we provide preliminaries on DP in Section~\ref{sec:prel:dp}, MI in Section~\ref{sec:prel:mi}, HTC in Section~\ref{sec:htc} and lastly HTC specific machine learning concepts in Section~\ref{sec:embeddings}. Throughout this paper we will use the abbreviations and variables denoted in Table~\ref{tab:Notation}.

\begin{table}[bt!]
	\footnotesize
	\caption{List of acronyms.}
	\label{tab:Notation}
	\centering
	\begin{tabularx}{.48\textwidth}{ >{\bfseries }lX}
        ANN & Artificial Neural Network \\
        AUC & Area under the ROC Curve \\
        BERT & Bidirectional Encoder Representations from Transformers \\
        BoW & Bag of Words \\
        CNN & Convolutional Neural Network \\
        DAG & Directed Acyclic Graph \\
        DP & Differential Privacy \\
        FPR & False Positive Rate \\
        HTC & Hierarchical Text Classification \\
        LCA & Lowest Common Ancestor \\
        LCL & Local Classifier per Level \\
        LCPN & Local Classifier per Parent Node \\
        LCN & Local Classifier per Node \\
        MI & Membership Inference \\
        NLP & Natural Language Processing \\
        RCV1 & Reuters Corpus Volume 1 \\
        RDP & Rényi Differential Privacy \\
        RNN & Recurrent Neural Network \\
        ROC & Receiver Operating Characteristic \\
        TPR & True Positive Rate \\
	\end{tabularx}
\end{table}

\label{sec:preliminaries}
\subsection{Differential Privacy}
\label{sec:prel:dp}
In DP~\cite{dworkDifferentialPrivacy2006} a statistical aggregation function $f(\cdot)$ is evaluated over a dataset $\mathcal{D}$, and the result is perturbed before being provided to the data analyst. By means of perturbation DP prevents an adversary with arbitrary auxiliary knowledge on all but one participant in a dataset $\mathcal{D}$ from confidently deciding whether $f(\cdot)$ was evaluated on $\mathcal{D}$, or some neighboring dataset $\mathcal{D'}$ differing in one element.
Assuming that every participant in $\mathcal{D}$ is represented by a single record $d\in\mathcal{D}$, privacy is intuitively provided to any individual. The perturbation strength is steered by setting privacy parameters $(\epsilon,\delta)$, and small privacy parameters will result in high perturbation. A formal definition of DP is provided in Definition~\ref{def:dp}.

\begin{definition}[$(\epsilon,\delta)$-DP~\cite{dworkAlgorithmicFoundationsDifferential2013}]
\label{def:dp}
A randomized mechanism $\mathcal{M}$ on a query function $f$ satisfies $(\epsilon,\delta)$-DP for $\delta>0$ if, for all pairs of neighboring databases $\mathcal{D},\mathcal{D'}$ and for all outputs $\mathcal{O}\subseteq$ \emph{range}($\mathcal{M}$),
\begin{equation}
    \label{eq:DP}
\Pr[\mathcal{M}(\mathcal{D})\in \mathcal{O}] \leq e^{\epsilon} \Pr[\mathcal{M}(\mathcal{D'})\in \mathcal{O}] + \delta.
\end{equation}
\end{definition}

DP is enforced by mechanisms. Mechanisms for numerical data perturb the original query value $f(\mathcal{D})$ by adding numerical noise. DP mechanisms need to add noise scaled to the \textit{global sensitivity}. Global sensitivity is formally defined in Definition~\ref{def:gs}. 

\begin{definition}[Global $\ell_1$-sensitivity~\cite{dworkDifferentialPrivacy2006}]
\label{def:gs}
Let $\mathcal{D}$ and $\mathcal{D'}$ be neighboring databases. The global $\ell_1$-sensitivity of a function $f$, denoted by $\Delta f$, is defined as
	\begin{equation*}
	\Delta f = \max_{\forall\mathcal{D},\mathcal{D'}}\|f(\mathcal{D}) - f(\mathcal{D'})\|_1.
	\end{equation*}
	\label{def:gs_eq}
\end{definition}

In this work, we use the Gaussian mechanism for gradient-perturbation to perturb the Adam optimizer for ANN training \footnote{The Tensorflow privacy package was used throughout this work: \url{https://github.com/tensorflow/privacy}.} as suggested by Abadi et al.~\cite{abadiDeepLearningDifferential2016}. For simplicity, we shall refer to the perturbed Adam optimizer as DP-Adam. A DP optimizer for ANN training, such as DP-Adam, uses a  randomized mechanism $\mathcal{M}_{nn}$. The optimizer updates the weight coefficients $\theta_t$ of an ANN per training step $t\in\{1,\ldots,T\}$ with $\theta_t \leftarrow \theta_{t-1}-\alpha(\tilde g)$, where $\tilde g~=~\mathcal{M}_{nn}(\partial loss / \partial \theta_{t-1})$ denotes a perturbed gradient. $\alpha$ is a scaling function on $\tilde g$ to compute an update (i.e., learning rate).  After $T$ steps, DP-Adam outputs a DP weight matrix $\theta$ that is used by the ANN prediction function. In case of DP-Adam, $\cali{M}_{nn}$ is a Gaussian mechanism as specified in Theorem~\ref{thm:gauss}.
\newline
\begin{theorem}[Gaussian Mechanism~\cite{dworkAlgorithmicFoundationsDifferential2013}]
\label{thm:gauss}~Let $\epsilon\in(0,1)$ be arbitrary. For $c^2>2ln(\frac{1.25}{\delta})$, the Gaussian mechanism with parameter $\sigma\ge c\frac{\Delta f}{\epsilon}$ satisfies $(\epsilon,\delta)$-DP, when adding noise scaled to the Normal distribution $\mathcal{N}(0,\sigma^2)$.
	\label{def:dp:gauss}
\end{theorem}

DP-Adam bounds the sensitivity of the computed gradients by a clipping norm $\mathcal{C}$, based on which the gradients are clipped before perturbation.
Since weight updates are performed iteratively during training, a composition of mechanism executions is required until the training step $T$ is reached and the final private weights $\theta$ are obtained.
We use R{\'e}nyi DP as suggested by Mironov~\cite{mironovRenyiDifferentialPrivacy2017} to calculate the tight, overall privacy guarantee $\eps$ under composition. $(\alpha,\eps_{RDP})-$R{\'e}nyi DP (RDP), with $\alpha >1$ quantifies the difference in distributions $\cali{M}(\cali{D}), \cali{M}(\cali{D'})$ by their R{\'e}nyi divergence~\cite{vanErven2010}. For a sequence of $T$ mechanism executions each providing ($\alpha$, $\eps_{RDP, i}$)-RDP, the privacy guarantee composes to ($\alpha$, $\sum_{i}\eps_{RDP, i}$)-RDP. The ($\alpha$, $\eps_{RDP}$)-RDP guarantee converts to $(\eps_{RDP}-\frac{\ln\dlt}{\alpha-1},\dlt)$-DP. The Gaussian mechanism is calibrated to RDP by:

\begin{equation}\label{eq:gaussrdp}
    \eps_{RDP} = \alpha \cdot \Delta f^2/2\sigma^2
\end{equation}

\subsection{Membership Inference}
\label{sec:prel:mi}
Membership Inference attacks strive for identifying the presence or absence of individual records in the training data of a machine learning model. Throughout this paper we refer to the trained machine learning model as \textit{target model} and the data owner's training data as $\mathcal{D}_{target}^{train}$. We solely consider ANNs as target models in this paper. ANNs are structured in layers of neurons that are connected by weights. We denote the weights between a layer $l$ and its preceding layer $l-1$ as $w^{(l)}$. The output of the $l$-th layer is denoted as $o^{(l)}$. The ANN's final output is the output of the last layer.

This paper builds upon the white-box MI attack against ANNs proposed by Nasr et al.~\cite{nasrComprehensivePrivacyAnalysis2018}. Essentially, the white-box MI attack assumes an honest-but-curious adversary with access to the target model weight matrix $w^{(l)}$. The white-box MI adversary leverages this knowledge to calculate attack features for each record $(x, y)$ in the form of layer outputs $o^{(l)}(x;W)$, losses $L(o(x;w),y)$, and  gradients $\frac{\partial L}{\partial w^{(l)}}$.

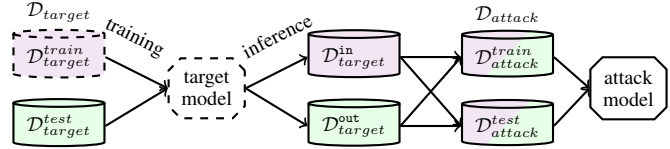
\begin{figure}
	\centering
	\begin{tikzpicture}[thick,scale=0.85, every node/.style={transform shape}]
	\footnotesize
	\node[] (aa) {};
	\node[draw,cylinder,shape border rotate=90, minimum width=4.5em, dashed, above= 0.25ex of aa, shape aspect=.10, fill=violet!10] (a) {$\mathcal{D}_{target}^{train}$};
	\node[minimum width=4em, above=0ex of a] (uu) {$\mathcal{D}_{target}$};
	\node[draw,cylinder,shape border rotate=90, minimum width=4.5em, below= 0.25ex of aa, shape aspect=.10, fill=green!10] (b) {$\mathcal{D}_{target}^{test}$};
	\node[draw,cylinder,shape border rotate=90, minimum width=4.5em, right= 10em of a, shape aspect=.10, fill=violet!10] (c) {$\mathcal{D}_{target}^{\var{in}}$};
	\node[draw,cylinder,shape border rotate=90, minimum width=4.5em, right= 10em of b, shape aspect=.10, fill=green!10] (d) {$\mathcal{D}_{target}^{\var{out}}$};
	\node[draw,cylinder,shape border rotate=90, minimum width=4.5em, right= 3em of c, shape aspect=.10, double color fill={violet!10}{green!10}, shading angle=45] (e) {$\mathcal{D}_{attack}^{train}$};
	\node[draw,cylinder,shape border rotate=90, minimum width=4.5em, right= 3em of d, shape aspect=.10, double color fill={violet!10}{green!10}, shading angle=45] (f) {$\mathcal{D}_{attack}^{test}$};
	\node[minimum width=4em, above=0ex of e] (uu) {$\mathcal{D}_{attack}$};
	\node[draw,chamfered rectangle, dashed, align=center, right=5em of aa] (z) {target \\ model};
	\node[draw,chamfered rectangle, align=center, right=17em of z] (y) {attack \\ model};

	\draw [->] (a.east) --  node [near start, above=1.5ex,sloped]{\small training}(z.west);
	\draw [->] (b.east) --  (z.west);
	
	\draw [->] (z.east) -- node [near end, above=2ex,sloped]{\small inference}(c.west);
	\draw [->] (z.east) -- (d.west);
	
	\draw [->] (c.east) -- (e.west);
	\draw [->] (c.east) -- (f.west);
	
	\draw [->] (d.east) -- (e.west);
	\draw [->] (d.east) -- (f.west);
	
	\draw [->] (e.east) -- (y.west);
	\draw [->] (f.east) -- (y.west);
	
	\end{tikzpicture}
	\caption{White-box MI with attack features. 
	DP perturbation is applied on the target model training (dashed). The data that was used by the data owner during target-model training is colored: training (violet) and validation (green).} 
	\label{fig:wb_mia}
\end{figure}

With the aforementioned data the white-box MI adversary trains a binary classifier, the \textit{attack model}. The attack model allows to classify records into members and non-members w.r.t.~the target model and training dataset. The adversary is assumed to know a portion of the training and test data $\mathcal{D}_{target}^{train}$ and $\mathcal{D}_{target}^{test}$, and generates features for training the attack model by passing the known records repeatedly through the trained target model. 
Nasr et al.~\cite{nasrComprehensivePrivacyAnalysis2018} assumed the portion of known records at 50\% and we follow this assumption to allow comparison. The performance metrics of an MI attack model are typically evaluated on a balanced dataset including members (target model training data) and an equal number of non-members (target model test data). An illustration of the data preparation for the white-box MI attack and its evaluation is shown in Figure~\ref{fig:wb_mia}.

\subsection{Hierarchical Text Classification}
\label{sec:htc}

The task of \emph{text classification} consists in categorizing texts into a pre-defined set of categories that do not have any hierarchical structure. Text classification is crucial in NLP, \cite{manningFoundationsStatisticalNatural}, a subfield of linguistics, computer science, and artificial intelligence
that is concerned with how computers can be programmed to process and analyze natural language data.

On the other hand, HTC addresses the task of classifying a text document into a \emph{hierarchy of classes and sub-classes}. Text classification can therefore be regarded as a special case of HTC with only one hierarchy level and no sub-categories. Hierarchical classification problems can be categorized on the basis of their \emph{hierarchy structure}, \emph{label type} and \emph{label depth}. In this work, we shall consider \textit{tree} hierarchies with \textit{single} \textit{partial-depth} labels, which means that every text shall be assigned to a single label that can be any node in a tree hierarchy.

In many circumstances, hierarchical classification may be desirable as categorizing documents into varying levels of abstraction better fits the nature of certain applications, e.g., product categorization or support-ticket classification. 
Also, as it has been consistently shown by numerous cognitive studies~\cite{murphy2004big}, people tend to favor categorization at different levels of abstraction.

Formally, HTC comprises a collection of text documents $x_1,\ldots,x_j\in \mathbb{X}$, where $\mathbb{X}$ is a document space; and a fixed set of classes $\mathbb{Y} = \{ y_1,y_2,\ldots,y_k \}$ belonging to some hierarchy. Given a training set of labeled documents $(x_1,y_1),\ldots,(x_n,y_n)$ on a hierarchy, where $(x_i,y_i) \in \mathbb{X} \times \mathbb{Y}$, we wish to learn a classifier or classification function $\gamma $ that maps documents to classes,
\begin{displaymath}
\gamma: \mathbb{X} \rightarrow \mathbb{Y},
\end{displaymath}
\noindent
so that each text document is only assigned to a single label, and a certain utility metric (see Sec.~\ref{sec:methodology:utility}) is maximized.

\subsection{Embeddings}
\label{sec:embeddings}
ANNs employ \emph{embeddings} in the first layer to capture the meaning of each token. An embedding is a token's vector representation of length $n$, that embeds the token into an $n$-dimensional vector space~\cite{goyalDeepLearningNatural2018}. Although embeddings have the ability to map semantically similar tokens to the same region in the vector space, a common shortcoming is that a word is always assigned to the same vector, ignoring previous and subsequent words (e.g., see  Word2Vec~\cite{mikolovEfficientEstimationWord2013,penningtonGloveGlobalVectors2014}). 

Word embeddings enable transfer learning, meaning that they can be trained on large unlabeled text corpora and afterwards be used in NLP systems for various tasks. In the case of ANNs, this is achieved by pre-initializing the first layer of an ANN with the pre-trained word embeddings. 
\section{Quantifying Utility and Privacy in HTC}
\label{sec:methodology}

\begin{figure*}
\captionsetup{justification=centering}
    \centering
	\begin{subfigure}{0.5\linewidth}
        \centering
        \includegraphics[width=\textwidth]{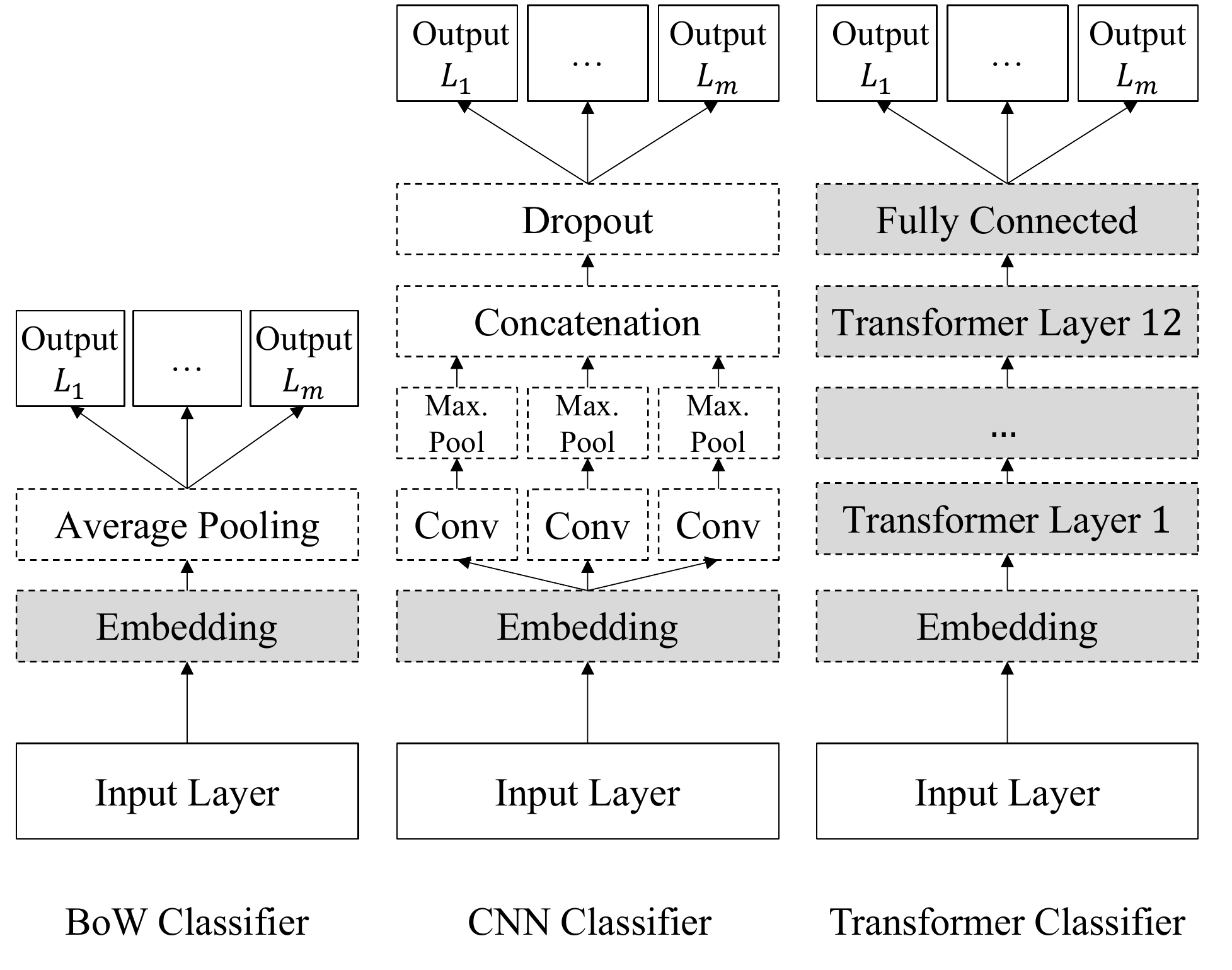}
        \caption{Architecture of the HTC models used in the experiments. Pre-trained layers are marked \textit{grey}.}
        \label{fig:htc-models}
    \end{subfigure}%
	\begin{subfigure}{0.5\linewidth}
        \centering
        \includegraphics[width=\textwidth]{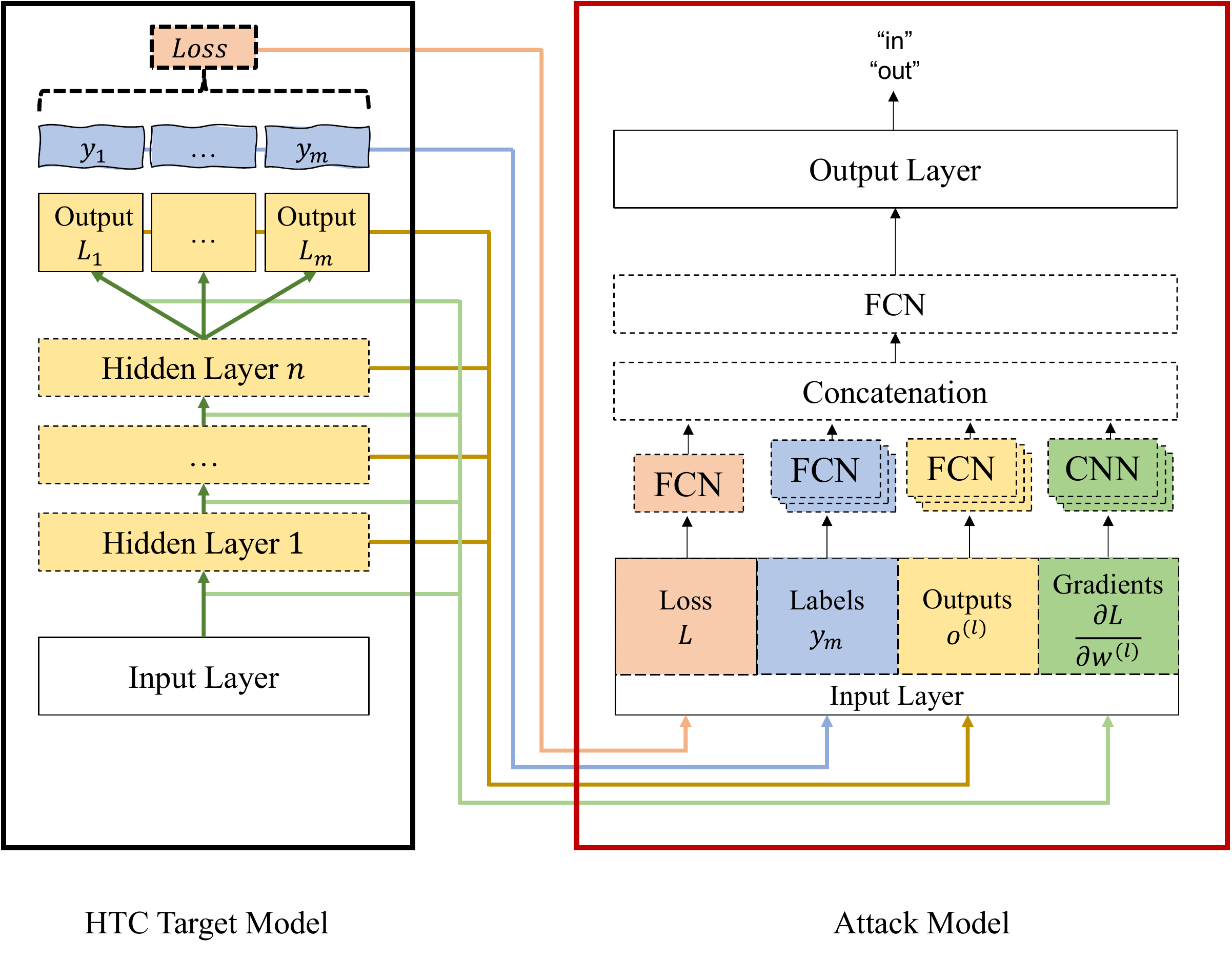}
        \caption{Attack model architecture and observed attack features from an HTC target model.}
        \label{fig:htc-models:attack}
    \end{subfigure}
    \label{fig:models}
    \caption{Architecture of target models and attack model.}
\end{figure*}

This section describes our methods for quantifying and comparing the privacy-utility trade-off in HTC for three relevant model architectures under several utility and MI metrics.

\subsection{HTC Model Architectures}

Architectures for HTC include training a single classifier predicting classes in the flattened hierarchy (\textit{flat}), training multiple classifiers predicting classes for a given level or node (\textit{local}) and training a single classifier that respects the class hierarchy (\textit{global}).

We chose a global HTC approach that features a single ANN with one output layer per level $L_n$ in the tree hierarchy of the given HTC task. Each output layer is a fully connected layer with a \textit{softmax} activation  which uses the output of the last hidden layer to predict the class for the input text on $L_n$. The architecture thus consists of only one model and does not ignore the class hierarchy. This yields two main advantages. First, our HTC classifier exhibits a  reduced training time and therefore also a lower overall privacy cost in comparison to local HTC approaches. Secondly, the data analyst can still retrieve the prediction probabilities per level in contrast to flat HTC approaches that only provide prediction probabilities of the individual nodes~\cite{Babbar13NIPS}.

Our HTC approach, however, has the disadvantage that a post-processing step is needed to obtain predictions consistent with the hierarchy, since each output layer makes predictions independently for its hierarchy level. We show an example in Figure~\ref{fig:lcl-incconsistencies} where the prediction for $L_2$ does not coincide with the prediction for $L_1$ and $L_3$, resulting in an undefined assignment. We suggest to resolve such inconsistencies by multiplying the softmax probabilities along the path from the root to each possible node. After comparing the multiplied probabilities, we output the path with the highest probability as prediction, leading to only consistent predictions.

Even though we defined the output layer architecture for our HTC beforehand, multiple options for the architecture of the input and intermediary layers exist. We consider architectures widely used in state-of-the-art text classification as the basis for formulating a DP hierarchical text classifier. In the sequel, we briefly describe the three architectures employed in our methodology.

\textbf{Bag-of-Words.} BoW models represent ANNs that ignore the word order within a text. BoW models comprise a single embedding layer in which the word vectors are added or averaged, and which is followed by one or more feed-forward layers with a softmax activation in the last layer. The BoW classifier used in this paper is based on the architecture of \textit{fastText}~\cite{joulinBagTricksEfficient2016}, which achieves high accuracy and is computationally efficient due to the usage of a single feed-forward layer for classification. Each token is first embedded and the mean of all embeddings before the output layer is computed afterwards. Prior to training, the embedding layer is initialized with the widely used GloVe embeddings\footnote{\url{https://nlp.stanford.edu/projects/glove/}} that are pre-trained on the \enquote{Wikipedia 2014} and \enquote{Gigaword~5} corpora~\cite{penningtonGloveGlobalVectors2014, kowsariTextClassificationAlgorithms2019}. For training we use the Adam optimizer. The classifier architecture is visualized in Figure~\ref{fig:htc-models}.

\textbf{Convolutional Neural Networks.} A CNN contains one or more convolutional layers that convolve the input with a filter of a given width to extract and detect patterns. Although CNNs are widely used for detecting patterns in images, CNNs have also been effectively used for detecting patterns in a text~\cite{kalchbrenner-etal-2014-convolutional, collobert2011natural} and do not ignore the word order. In this paper we use the original CNN classifier for text classification proposed by Kim~\cite{kimConvolutionalNeuralNetworks2014}. Their architecture first applies three convolutional layers to the embeddings, which are then concatenated and passed through a dropout layer to foster generalization. The architecture of the resulting HTC model is provided in Figure~\ref{fig:htc-models}. For training, we use the Adam optimizer and Glove embeddings. The training hyperparameters are taken from Kim~\cite{kimConvolutionalNeuralNetworks2014}: filter sizes of $3$, $4$, and $5$ for the three convolutional blocks with $100$ filters each and a dropout probability of $p_{do}=0.5$.

\textbf{Transformer Networks.} Transformer layers are ANN layers that are especially suited for processing longer texts due to a mechanism called \textit{self-attention}. Due to self-attention, a single Transformer layer can relate all tokens of a text to each other~\cite{vaswaniAttentionAllYou2017}. In contrast, CNNs require multiple convolutional layers to relate the information between two arbitrary tokens in a text. The transformer classifier we use in this paper is the BERT model as formulated by Devlin et al.~\cite{devlinBERTPretrainingDeep2018}.
BERT comprises twelve transformer layers, each consisting of two sub-layers. To lower the computational effort that is needed for training BERT we follow related work and initialize the BERT layers with pre-trained weights~\cite{devlinBERTPretrainingDeep2018, wolfHuggingFaceTransformersStateoftheart2020}. During training, we employ the Adam optimizer and a dropout probability of $p_{do}=0.1$. The BERT HTC architecture is illustrated in Figure~\ref{fig:htc-models}. 

\subsection{Utility Metrics}

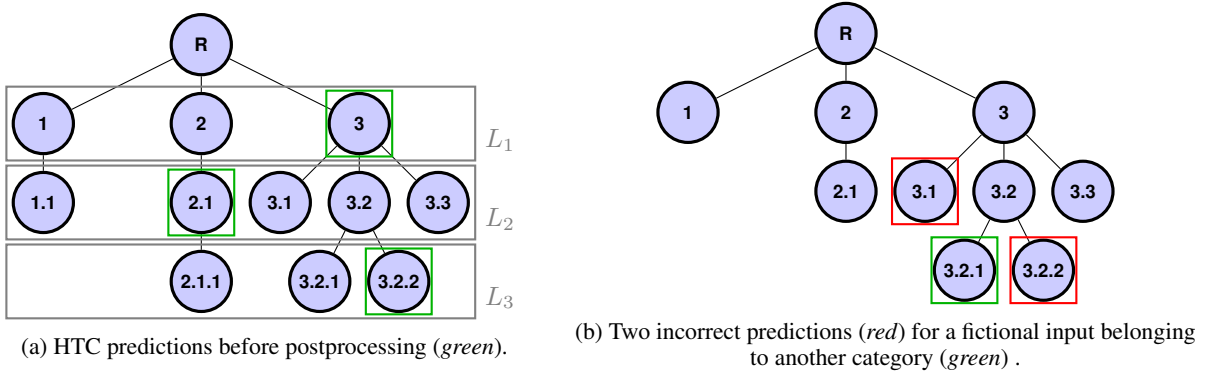
\begin{figure*}[ht!]
	\centering
    \captionsetup{justification=centering}
	\begin{subfigure}{0.5\linewidth}
    \centering
    \begin{tikzpicture}[auto,
        scale=0.7,
        level 1/.style={sibling distance=3cm},
        level 2/.style={sibling distance=15mm},
        level 3/.style={sibling distance=15mm},
        every node/.style = {circle, align=center, fill=blue!20, draw=black, very thick,
            minimum size = 8mm, font=\scriptsize\bf\sffamily,
            inner sep=0
        }]
        \node {R}
            child { node {1} 
                child { node (W) {1.1} } }
            child { node {2} 
                child { node (P2) {2.1} 
                    child { node {2.1.1} } } }
            child { node (P1) {3}
                child { node {3.1} }
                child { node {3.2}
                    child { node {3.2.1} }
                    child { node (P3) {3.2.2} } }
                child { node (E) {3.3} } };
                
        \tikzset{every node/.style={draw=none, fill=none}}
              
        \draw[black!50,thick]
            let
                \p1=(W.west),
                \p2=(P1.north),
                \p3=(E.east),
                \p4=(P1.south)
            in
                ($(\x1,\y2)+(-0.1,0.1)$) rectangle ($(\x3, \y4)+(0.1,-0.1)$) node[pos=1] {$L_1$};
                
        \draw[black!50,thick]
            let
                \p1=(W.west),
                \p2=(P2.north),
                \p3=(E.east),
                \p4=(P2.south)
            in
                ($(\x1,\y2)+(-0.1,0.1)$) rectangle ($(\x3, \y4)+(0.1,-0.1)$) node[pos=1] {$L_2$};
                
        \draw[black!50,thick]
            let
                \p1=(W.west),
                \p2=(P3.north),
                \p3=(E.east),
                \p4=(P3.south)
            in
                ($(\x1,\y2)+(-0.1,0.1)$) rectangle ($(\x3, \y4)+(0.1,-0.1)$) node[pos=1] {$L_3$};
                
        \draw[black!30!green,thick] ($(P1.north west)+(-0.2,0.2)$) rectangle ($(P1.south east)+(0.2,-0.2)$);
        \draw[black!30!green,thick] ($(P2.north west)+(-0.2,0.2)$) rectangle ($(P2.south east)+(0.2,-0.2)$);
        \draw[black!30!green,thick] ($(P3.north west)+(-0.2,0.2)$) rectangle ($(P3.south east)+(0.2,-0.2)$);
        
    \end{tikzpicture}
    \caption{HTC predictions before postprocessing (\textit{green}).}
    \label{fig:lcl-incconsistencies}
	\end{subfigure}%
	\begin{subfigure}{0.5\linewidth}
	    \centering
    \begin{tikzpicture}[auto,
        scale=0.7,
        level 1/.style={sibling distance=3cm},
        level 2/.style={sibling distance=15mm},
        level 3/.style={sibling distance=15mm},
        every node/.style = {circle, align=center, fill=blue!20, draw=black, very thick,
            minimum size = 8mm, font=\scriptsize\bf\sffamily,
            inner sep=0
        }]
        \node {R}
            child { node {1} }
            child { node {2} 
                child { node {2.1} } }
            child { node {3}
                child { node (P2) {3.1} }
                child { node {3.2}
                    child { node (T) {3.2.1} }
                    child { node (P1) {3.2.2} } }
                child { node {3.3} } };
        \draw[black!30!green,thick] ($(T.north west)+(-0.2,0.2)$)  rectangle ($(T.south east)+(0.2,-0.2)$);
        \draw[red,thick] ($(P1.north west)+(-0.2,0.2)$)  rectangle ($(P1.south east)+(0.2,-0.2)$);
        \draw[red,thick] ($(P2.north west)+(-0.2,0.2)$)  rectangle ($(P2.south east)+(0.2,-0.2)$);
    \end{tikzpicture}
    \caption{Two incorrect predictions (\textit{red}) for a fictional input belonging to another category (\textit{green}) .}
    \label{fig:example_tree}
	\end{subfigure}
	\caption{Visualization for obtaining and evaluating HTC predictions.}
\end{figure*}

\label{sec:methodology:utility}
There are two approaches for evaluating the utility provided of the described
model architectures, namely, flat and hierarchical evaluation metrics~\cite{sillaSurveyHierarchicalClassification2011, kosmopoulosEvaluationMeasuresHierarchical2015}. To illustrate the difference between flat and hierarchical classification metrics consider the tree hierarchy depicted in Figure~\ref{fig:example_tree}. Assume that the true category for a given test example $x$ is $3.2.1$ (\textit{green}) and that two different classifiers output $3.2.2$ and $3.1$ as the predicted categories (\textit{red}). When flat evaluation metrics are used, both systems are penalized equally since both predictions are counted as false negatives for the true category $3.2.1$. However, the second classifier's error is more severe since its prediction is in an unrelated sub-tree of node $3$, which is considered in hierarchical evaluation metrics.

In this work, we assess the utility of HTC based on a mix of flat and hierarchical metrics: \textit{accuracy}, the hierarchical and lowest common ancestor (LCA) \textit{F-measure}. We report the (flat) accuracy $Acc$ due to its wide use in machine learning. We calculate the \textit{hierarchical and LCA F-measure} since they are considered the state of the art in the field of HTC. The hierarchical $F$-measure $F_H$ is calculated from \emph{hierarchical precision} $P_H$ and \emph{hierarchical recall} $R_H$, and defined as follows~\cite{sillaSurveyHierarchicalClassification2011}:
\begin{align*}
        &P_H = \frac{\sum_i|Anc_i\cap\hat{Anc_i}|}{\sum_i|\hat{Anc_i}|},\\
        &R_H = \frac{\sum_i|Anc_i\cap\hat{Anc_i}|}{\sum_i|Anc_i|},\\
        &F_H = \frac{2P_HR_H}{P_H+R_H}.
\end{align*}
For a record $i$, $Anc_i$ is the set consisting of the \textit{true} class and all ancestors (except the root). Analogously, $\hat{Anc_i}$ is the set consisting of the \textit{predicted} class and all ancestors (except the root). In Figure~\ref{fig:example_tree} $Anc_i = \{3.2.1, 3.2, 3\}$ for the true class $3.2.1$. $\hat{Anc_i}$ is $\{3.2.2, 3.2, 3\}$ and $\{3.1, 3\}$ with $R_H = \frac{2}{3}$ and $R_H = \frac{1}{3}$, respectively. 

The LCA metrics  $P_{LCA}, R_{LCA}, F_{LCA}$ are differing to the previous hierarchical metrics only by not considering the nodes above the lowest common ancestor in $Anc_i$, and thus void overpenalization of errors for nodes with many ancestors~\cite{kosmopoulosEvaluationMeasuresHierarchical2015}. Again, in the example shown in Figure~\ref{fig:example_tree}, for the prediction $3.2.2$ we would have $LCA = 3.2$ with $Anc_i = \{3.2.1, 3.2\}$ and $\hat{Anc_i} = \{3.2.2, 3.2\}$ and therefore $R_{LCA} = \frac{1}{2}$. For the prediction $3.1$ we would have $LCA = 3$ with $Anc_i = \{3.2.1, 3.2, 3\}$ and $\hat{Anc_i} = \{3.1, 3\}$ and therefore $R_{LCA} = \frac{1}{3}$.

\subsection{Privacy Metrics and Bounds}
DP formulates a privacy bound on the ratio of probability distributions around $D$ and $D'$ resulting from a mechanism. The privacy bound holds for an adversary with auxiliary knowledge of up to all but one records in the dataset~\cite{Lee2012,BEGKK21}. Yeom et al.~\cite{yeomPrivacyRiskMachine2018} demonstrate that the privacy bound can be transformed into an upper bound on the membership advantage of an MI adversary. Membership advantage is calculated as follows from the True Positive Rate (TPR) and the False Positive Rate (FPR)~\cite{FAWCETT2006861}:
\begin{equation*}
    Adv = TPR - FPR.
\end{equation*}
The upper bound is:
\begin{equation}
\mathit{Adv}\leq e^\epsilon - 1.
\end{equation}
Whether the resulting membership advantage upper bound is reached, i.e.~the empirically observed $Adv$ matches the upper bound, depends on whether the sensitivity of the training data during model training matches the assumed global sensitivity (i.e., clipping norm $\mathcal{C}$; cf.~Theorem~\ref{thm:gauss})~\cite{Nissim2007}. The gap between the lower and upper bound can be validated by implementing an MI adversary~\cite{jayaramanEvaluatingDifferentiallyPrivate2019,bernauAssessingDifferentiallyPrivate2020}. 

Figure \ref{fig:htc-models:attack} visualizes the architecture of our implemented MI adversary's attack model, which is based on the attack model of Nasr et al.~\cite{nasrComprehensivePrivacyAnalysis2018}. We mainly extended their attack model to accept multiple labels, one per hierarchy level. The remaining components are unchanged. The attack model itself is represented by an ANN that learns to discriminate between training data and test data based on the attack features (e.g., losses).

In addition to $Adv$ we also quantify the area under the Receiver-Operating-Characteristic-Curve (AUC) of the attack model. The AUC is also providing insights on the MI attack performance w.r.t.~addressing members and non-members. The AUC is a general performance metric for evaluation of binary classifiers such as the MI attack model~\cite{FAWCETT2006861}.

\section{Reference datasets}
\label{sec:data}

We consider three real-world datasets: the BestBuy dataset, which represents a consumer product hierarchical classification task; the Reuters dataset, which contains news articles; and the DBPedia dataset with Wikipedia excerpts. The datasets have varying text lengths (34 to 212 words), differ in the number of overall data (51,000 to 800,000 records) and have also been used in related work on HTC without DP and MI.
\\
\\
\textbf{BestBuy.} The BestBuy dataset\footnote{\url{https://github.com/BestBuyAPIs/open-data-set}} contains $51,646$ unique products, each consisting of categorical features (e.g., SKU, type, manufacturer), numerical features such as price, textual features (e.g., name, description) and URLs, that are composed of one or more of the aforementioned features. Training a differentially private HTC model on datasets like BestBuy can therefore prevent leaking sensitive information about individual products of a company. 

In our experiments, we concatenate the features \enquote{name}, \enquote{manufacturer} and \enquote{description} to a single string and ignore the other features for classification. This selection is based on empirically observed superior classification accuracy. On average, the resulting concatenated texts have a length of 34 words. Additionally, every product holds a special feature called \enquote{category} assigning the product to a \textit{single}, \textit{partial-depth} class label in the BestBuy product hierarchy. 

The BestBuy product hierarchy is a \textit{tree} and consists of seven levels, each with a different number of classes, as shown in Table~\ref{tab:dataset-hierarchy}. As can be seen, level $L_4$ has the most classes. Also, we can see that even on the first level, not all of the existing $51,646$ products are assigned to a class. Particularly, we found that 256 products ($0.50$\%) are assigned to classes not contained in the BestBuy product hierarchy. We removed these products as the assigned classes did not fit into the given product hierarchy (e.g., \enquote{Other Product Categories} or \enquote{In-Store Only}). Furthermore, not every product is assigned to a class on every level, meaning the most specific class of many products is on a lower level than $L_7$. In our experiments, we only make use of the first three hierarchy levels. We decided to do so due to the long tail characteristic of the dataset. Thus, the predictions of our classifiers are less specific than potentially possible, but more robust due to a higher number of training examples in comparison to fine grained training for all hierarchies. 10\% of the overall data was used for testing.

All datasets have been \emph{tokenized}, which means that the text has to be split up into a sequence of smaller units called \emph{tokens}. A natural tokenization technique is splitting the text into a sequence of words, so that each token represents a word. After tokenization, the token sequence is converted into an integer sequence since ANNs only take numbers as input. During the conversion, a vocabulary is created that maps each token to a unique integer so that the same token is always converted into the exact same integer. The size of the vocabulary then represents the number of unique tokens in the text.
\begin{table}[ht!]
\centering
\begin{tabular}{c|c|c|c}
Hierarchy Level & Dataset & Classes & Assigned products \\ \hline
\multirow{3}{*}{Level $L_1$} & BestBuy & $19$ & $51,390$ \\ \cline{2-4} 
 & DBPedia & $9$ & $337,739$ \\ \cline{2-4} 
 & Reuters & $4$ & $804,427$ \\ \hline
\multirow{3}{*}{Level $L_2$} & BestBuy & $164$ & $50,837$ \\ \cline{2-4} 
 & DBPedia & $70$ & $337,739$ \\ \cline{2-4} 
 & Reuters & $55$ & $779,714$ \\ \hline
\multirow{3}{*}{Level $L_3$} & BestBuy & 612 & $44,949$ \\ \cline{2-4} 
 & DBPedia & 219 & $337,739$ \\ \cline{2-4} 
 & Reuters & 43 & $406,961$ \\ \hline
Level $L_4$ & BestBuy & $771$ & $26,138$ \\ \hline
Level $L_5$ & BestBuy & $198$ & $5,640$ \\ \hline
Level $L_6$ & BestBuy & $23$ & $346$ \\ \hline
Level $L_7$ & BestBuy & $1$ & $1$ \\
\end{tabular}
    \caption{Classes and assigned records per level per dataset.}
    \label{tab:dataset-hierarchy}
\end{table}
\\
\\
\textbf{Reuters Corpus Volume 1.} The \enquote{Reuters Corpus Volume 1} (RCV1) dataset\footnote{\url{https://trec.nist.gov/data/reuters/reuters.html}} is an archive of over $800,000$ manually categorized news articles~\cite{lewisRCV1NewBenchmark2004}. Per news article, a headline, text block and topic codes representing the classes in the hierarchy are provided. In our experiments we use the concatenation of headline and text block as input for the respective classifiers. The resulting texts have an average length of 237 words. Table~\ref{tab:dataset-hierarchy} shows the number of classes and assigned documents for each hierarchy level of the Reuters dataset. To ensure comparability with state of the art we follow the approach of Stein et al.~\cite{steinAnalysisHierarchicalText2019} and randomly assign $80,443$ texts to the test dataset and assign each news article to the least frequent topic code. This approach is based on the assumption that the least common topic code is the one that most specifically characterizes the document. A differentially private HTC model trained on a non-public article dataset such as RCV1 can prevent leaking sensitive information represented by individual articles.
\\
\\
\textbf{DBPedia.} DBPedia is a community project that extracts structured knowledge from Wikipedia and makes it freely available using linked data technologies~\cite{lehmannDBpediaLargescaleMultilingual2015}. The DBPedia dataset for HTC\footnote{\url{https://www.kaggle.com/danofer/dbpedia-classes}} is used as a reference dataset in many state-of-the-art publications~\cite{joulinBagTricksEfficient2016, yangXLNetGeneralizedAutoregressive2020, minaeeDeepLearningBased2020} on text classification. Overall, the dataset contains the introductions of $337,739$ Wikipedia articles, of which $240,942$ are pre-assigned to the training dataset and $60,794$ to the test dataset. Per article, the dataset contains a description of on average 102 words and three one-class label per hierarchy level ($L_1$, $L_2$, $L_3$). Table~\ref{tab:dataset-hierarchy} shows the number of classes on each level of the DBPedia dataset hierarchy and indicates that all texts are assigned to a class on all levels, which means that the labels are \textit{full depth}. Training a differentially private HTC model on a dataset like DBPedia prevents leaking information about participating institutions and people if the underlying encyclopedia is not public.

\section{Experimental Setup}
\label{sec:evaluation:setup}
For our experiments we split the datasets into training, validation and test data. Training data is used to learn the model parameters (i.e., weights), validation data to check the goodness of training hyperparameters and test data is used to assess generalization and real-world performance. Before target and attack model training, so called hyperparameters have to be set manually before training (e.g., learning rate, batch size). We used Bayesian hyperparameter optimization for all target model experiments to ensure that we found good hyperparameters that yield high accuracies on the respective models and data. Bayesian Optimization is more efficient than grid search since it considers past trials during the hyperparameter search. An overview of all hyperparameters, dataset size for training, validation and test, and the overall $\eps$ per training is provided in Table~\ref{tab:hyp}. For the attack model we reused the original hyperparameters of Nasr et al.~\cite{nasrComprehensivePrivacyAnalysis2018} which already performed well. 
The majority of experiments were conducted on EC2\footnote{\url{https://aws.amazon.com/ec2/}} GPU optimized instances of type \enquote{p3.8xlarge} with the \enquote{Deep Learning AMI} machine image, building on a Linux~$4.14$ kernel, Python~$3.6$ and TensorFlow~$2.2$.

\begin{table*}[t!]
\centering
\small
\caption{Hyperparameters and \eps~per model and dataset. The hyperparameters were set with Bayesian optimization.}
 \label{tab:hyp}
\begin{tabular}{c|c|c|c|c|c|c|c|c|c|c}
\hline
\multicolumn{2}{c|}{\multirow{2}{*}{}} & \multicolumn{3}{c|}{BestBuy} & \multicolumn{3}{c|}{Reuters} & \multicolumn{3}{c}{DBPedia} \\ \cline{3-11} 
\multicolumn{2}{c|}{} & BoW & CNN & Transformer & BoW & CNN & Transformer & BoW & CNN & Transformer \\ \hline
\multirow{2}{*}{learning rate} & Orig. & $0.001$ & $0.001$ & $0.005$ & $0.001$ & $0.001$ & $0.005$ & $0.001$ & $0.001$ & $0.005$ \\ \cline{2-11} 
 & DP & $0.01$ & $0.001$ & $0.015$ & $0.008$ & $0.001$ & $0.005$ & $0.016$ & $0.001$ & $0.01$ \\ \hline
\multirow{2}{*}{batch size} & Orig. & $32$ & $32$ & $32$ & $32$ & $32$ & $32$ & $32$ & $32$ & $32$ \\ \cline{2-11} 
 & DP & $64$ & $64$ & $64$ & $64$ & $64$ & $32$ & $64$ & $64$ & $32$ \\ \hline
$\mathcal{C}$ & DP & $0.19$ & $1.48$ & $2.07$ & $0.33$ & $6.28$ & $12.86$ & $0.03$ & $0.21$ & $1.6$ \\ \hline
microbatch size & DP & $1$ & $1$ & $1$ & $1$ & $1$ & $4$ & $1$ & $1$ & $4$ \\ \hline
\multirow{4}{*}{$\epsilon$} & $z=0.1$ & $30,253$ & $33,731$ & $5,902$ & $2,048$ & $1,091$ & $21,597$ & $4,317$ & $6,414$ & $24,741$ \\ \cline{2-11} 
 & $z=0.5$ & $11.5$ & $11.1$ & $6.58$ & $4.19$ & $4.11$ & $4.4$ & $5.1$ & $6.29$ & $4.88$ \\ \cline{2-11} 
 & $z=1.0$ & $1.51$ & $1.38$ & $1.04$ & $0.79$ & $0.79$ & $0.77$ & $0.87$ & $0.96$ & $0.81$ \\ \cline{2-11} 
 & $z=3.0$ & $0.26$ & $0.5$ & $0.29$ & $0.22$ & $0.22$ & $0.22$ & $0.2$ & $0.21$ & $0.21$ \\ \hline
\multicolumn{2}{c|}{Training records} & \multicolumn{3}{c|}{$41,625$} & \multicolumn{3}{c|}{$651,585$} & \multicolumn{3}{c}{$240,942$} \\ \hline
\multicolumn{2}{c|}{Validation records} & \multicolumn{3}{c|}{$4,626$} & \multicolumn{3}{c|}{$72,399$} & \multicolumn{3}{c}{$36,003$} \\ \hline
\multicolumn{2}{c|}{Test records} & \multicolumn{3}{c|}{$5,139$} & \multicolumn{3}{c|}{$80,443$} & \multicolumn{3}{c}{$60,794$} \\ \hline

\end{tabular}
\end{table*}

In all experiments we assume that the data owner would also want converging target models even when training with DP. Thus, all HTC models leverage early stopping with a patience of $3$ epochs to terminate the training process before overfitting. Furthermore, we set the DP parameter $C$ (i.e., the sensitivity $\Delta f$) in our experiments to the median of the norms of the unclipped gradients over the course of original training as suggested by Abadi et al.~\cite{abadiDeepLearningDifferential2016}. For all executions of the experiment, CDP noise is sampled from a Gaussian distribution (cf.~Definition~\ref{def:dp:gauss}) with scale $\sgm=\textit{noise multiplier}~z \times \textit{clipping norm}~\cali{C}$. According to McMahan et al.~\cite{mcmahanGeneralApproachAdding2019}, values of $z\approx 1$ will provide reasonable privacy guarantees. We evaluate increasing noise regimes per dataset by evaluating noise multipliers $z\in\{0.1, 0.5, 1.0, 3.0\}$ and calculate the resulting $\eps$~at a fixed $\dlt=\frac{1}{n}$. 

\section{Evaluation}
\label{sec:evaluation}


In this section, we first describe the experimental setup. Afterwards, we experimentally assess privacy and utility for the previously formulated HTC models and datasets\footnote{We publish all code and experiment scripts at \url{https://github.com/SAP-samples/security-research-dp-hierarchical-text}.}. Lastly, we present several experiment variations to illustrate the impact of different parameters.

\subsection{Empirical Privacy and Utility}
\label{sec:evaluation:results}

\noindent
A theoretic comparison of the CNN, BoW and Transformer models with respect to their robustness towards noise that is introduced by DP is only insightful to a limited extend, since their architectures and pre-training paradigms vary. However, in general the bias-variance trade-off for ANNs allows us to formulate high-level expectations. Simple ANNs will likely be prone to high bias and thus underfit in comparison to larger ANNs. Thus, the BoW model will potentially perform poorer on test data than the CNN or Transformer architecture, even in the presence of pre-training~\cite{ezen20}. In contrast, large ANNs will have high variance and thus require larger amounts of training data to generalize well. Thus, the Transformer model will likely perform poorer on small datasets. In general, the bias decreases and the variance increases with the ANN size~\cite{NMB+19}. In combination with DP, we expect high bias models such as the BoW to be less affected by the introduced noise. Additionally, Transformer models may be negatively affected by gradient explosion when using relu activation functions in combination with DP~\cite{papernot2020tempered}.

\begin{figure*}[ht!]
\captionsetup{justification=centering}
    \centering
        \begin{subfigure}{0.75\textwidth}
        \centering
        \includegraphics[width=\textwidth]{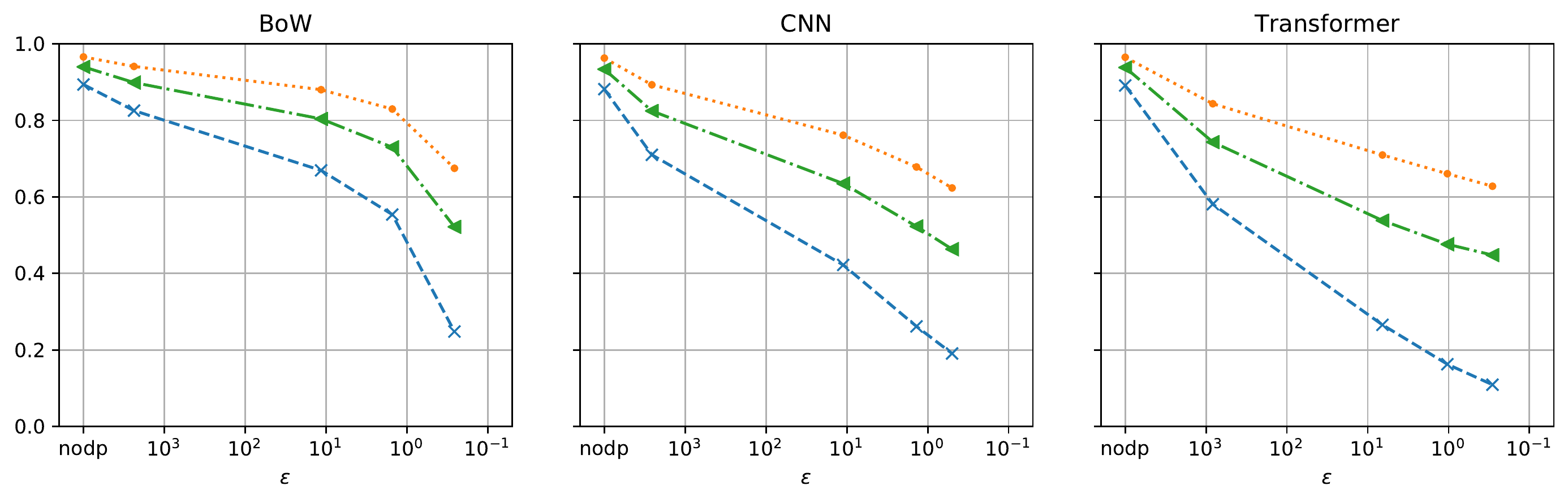}
        \caption{Target model test accuracy for BestBuy utility over $\epsilon$}
        \label{fig:bestbuy_cleaned_utility}
    \end{subfigure}
    \begin{subfigure}{0.75\textwidth}
        \centering
        \includegraphics[width=\textwidth]{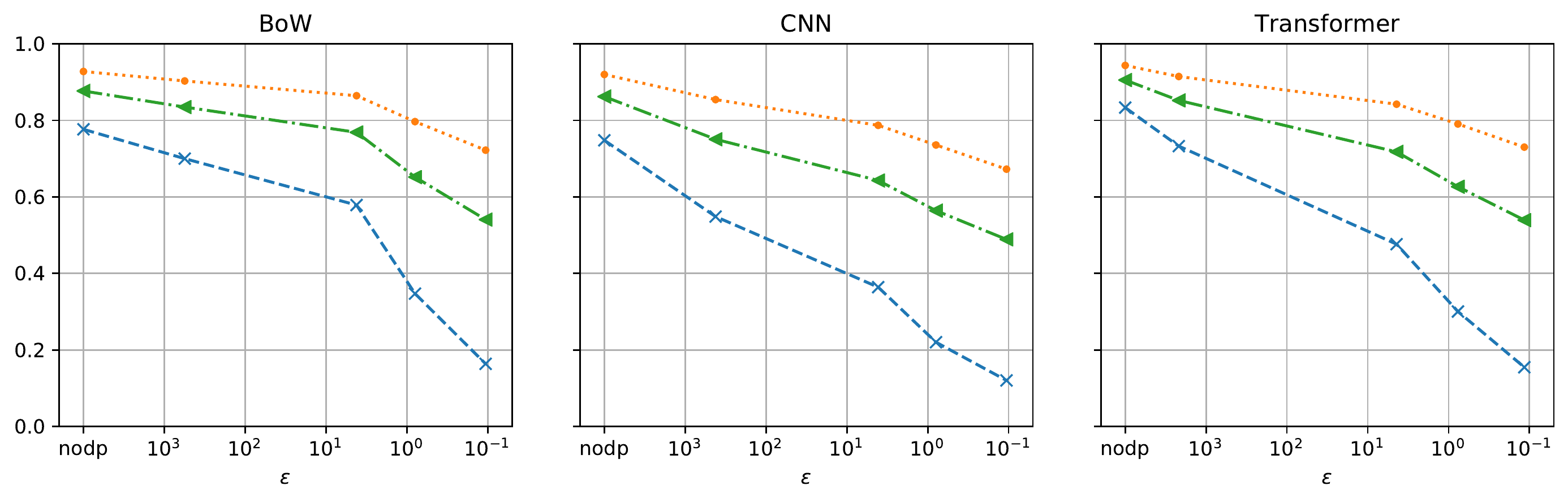}
        \caption{Target model test accuracy for Reuters utility over $\epsilon$}
        \label{fig:reuters_utility}
    \end{subfigure}
    \begin{subfigure}{0.75\textwidth}
        \centering
        \includegraphics[width=\textwidth]{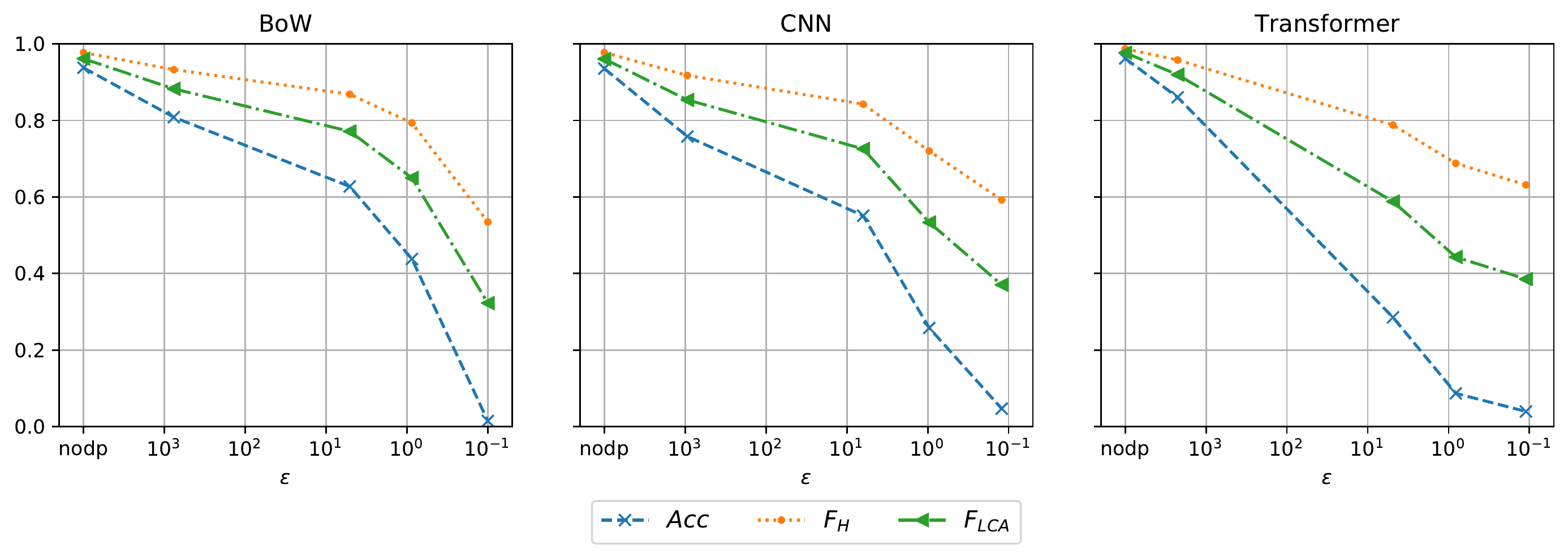}
        \caption{Target model test accuracy for DBPedia over $\epsilon$}
        \label{fig:dbpedia_utility}
    \end{subfigure}
    \caption{Target model test accuracy per dataset over $\epsilon$.}
    \label{fig:utility}
\end{figure*}

\begin{figure*}[ht!]
\captionsetup{justification=centering}
    \centering
    \begin{subfigure}{0.75\textwidth}
        \centering
        \includegraphics[width=\textwidth]{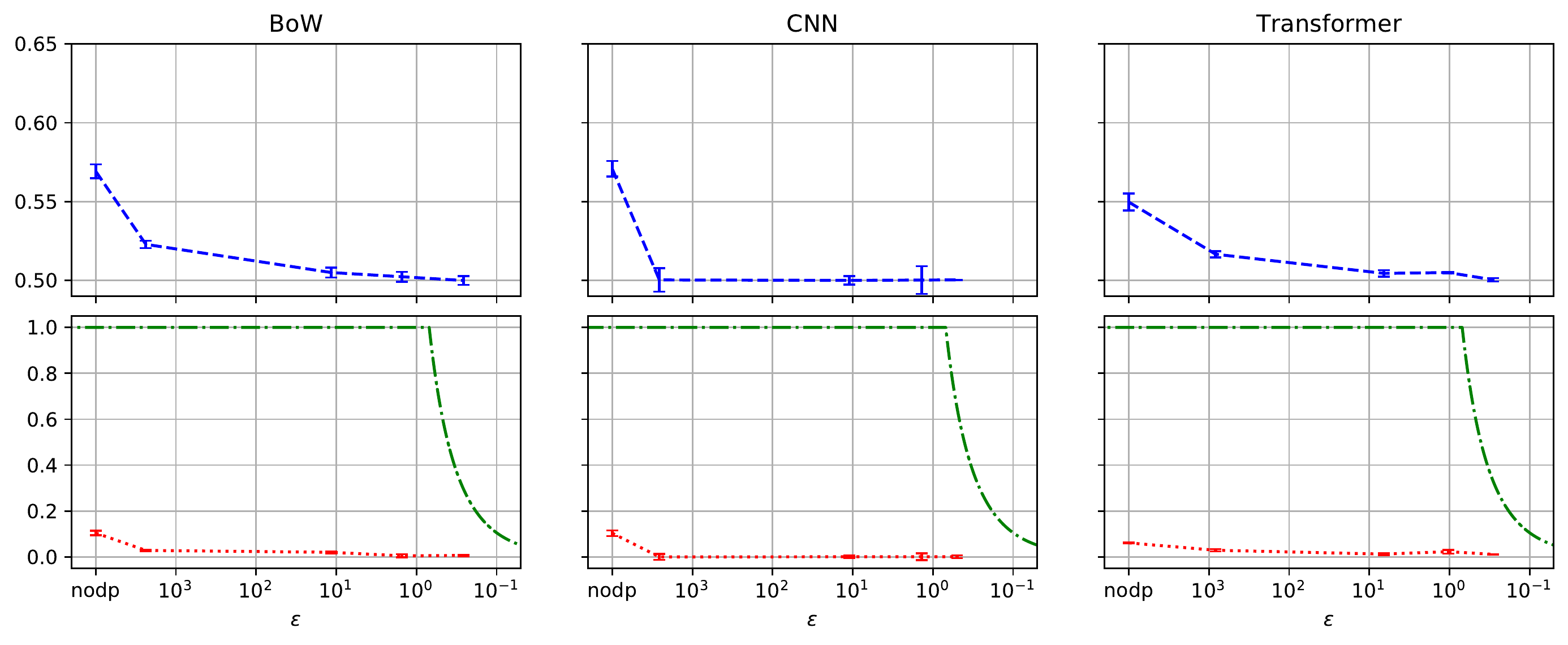}
        \caption{MI against BestBuy over $\eps$}
        \label{fig:bestbuy_cleaned_mi}
    \end{subfigure}
    \begin{subfigure}{0.75\textwidth}
        \centering
        \includegraphics[width=\textwidth]{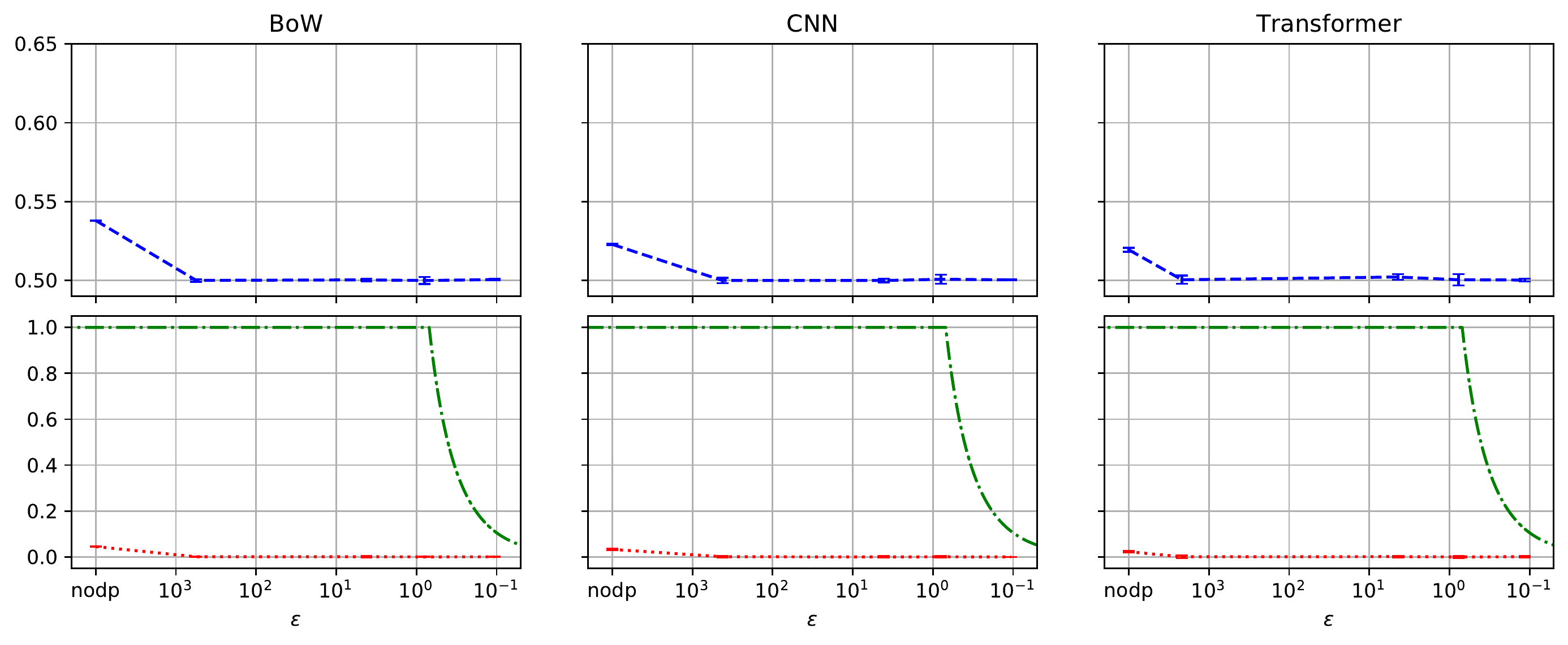}
        \caption{MI against Reuters over $\eps$}
        \label{fig:reuters_mi}
    \end{subfigure}
    \begin{subfigure}{0.75\textwidth}
        \centering
        \includegraphics[width=\textwidth]{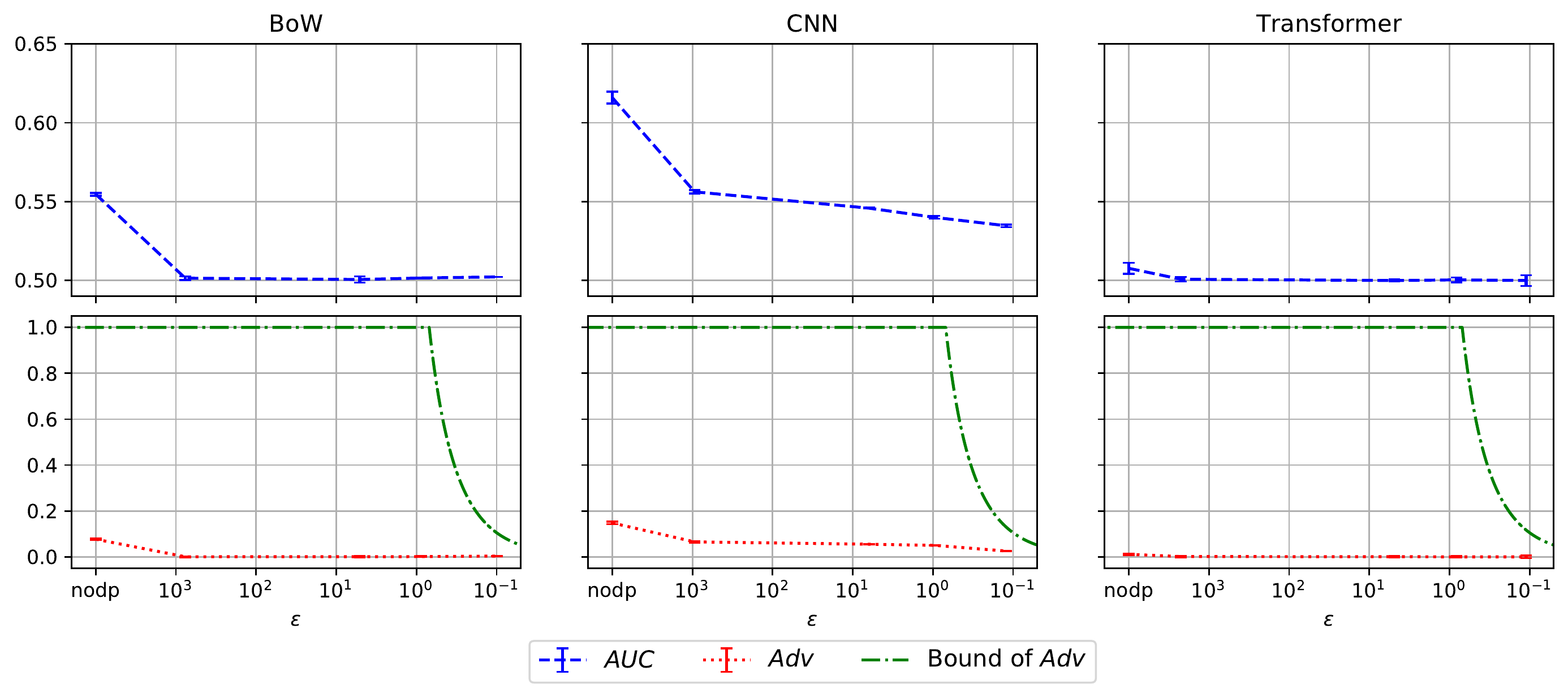}
        \caption{MI against DBPedia over $\eps$}
        \label{fig:dbpedia_mi}
    \end{subfigure}
    \caption{MI AUC, $Adv$ and Bound on MI $Adv$ per dataset $\epsilon$.}
    \label{fig:mi}
\end{figure*}

Figure~\ref{fig:utility} states the utility and Figure~\ref{fig:mi} the privacy scores over $\epsilon$ for the three datasets and model architectures. Furthermore, we additionally report the theoretical bound on $Adv$ by Yeom et al.~\cite{yeomPrivacyRiskMachine2018} to allow comparison of the theoretical and the empirical MI advantage. Notably, even if two classifiers were trained with the same noise multiplier $z$, they do not necessarily yield the same DP privacy parameter $\epsilon$ due to differing training epochs until convergence. All corresponding $\epsilon$ values per model and dataset were calculated for $\delta = \frac{1}{|\cali{D}_{target}^{train}|}$ per dataset and are stated in Table~\ref{tab:hyp}. 

As expected, the model utility and adversary's success consistently decrease with stronger DP parameters for all models and all datasets. Figure~\ref{fig:bestbuy_cleaned_utility} shows that 
for BestBuy the BoW model's utility is the most robust to the introduced noise, while the Transformer model's utility is most sensitive to the introduced noise. This observation becomes most evident when considering the flat accuracy $Acc$ (\textit{blue}), and is in line with our expectation for small datasets formulated at the beginning of this section. The hierarchical utility metrics $F_H$ and $F_{LCA}$ do not decrease as strongly as $Acc$, since they also account for partially correct predictions. Interestingly, for BestBuy, the CNN model's MI metrics in Figure~\ref{fig:bestbuy_cleaned_mi} already reach the baseline level at $\epsilon=33,731$ ($z=0.1$). The large $\epsilon$ points out that with respect to the upper bound a huge privacy loss is occurring (i.e., $e^\epsilon$) and the advantage should also be maximal (i.e., $e^\epsilon-1$~\cite{yeomPrivacyRiskMachine2018}). However, the empirical membership advantage lies far below this theoretical bound. In contrast to the CNN, the MI attack against the BoW and Transformer models is only reaching the baseline at $\epsilon=1$ and $\epsilon=1.5$, respectively.

The results for the Reuters dataset are provided in Figure~\ref{fig:reuters_utility} and \ref{fig:reuters_mi}. Compared to BestBuy, the decrease in model utility on Reuters is smaller for all three HTC models, which can be explained with a significantly higher amount of training examples and a smaller amount of hierarchical classes. The BoW classifier's utility is most robust to the addition of noise to the training process, yet closely followed by the Transformer model. However, the CNN model exhibits the most severe decrease in model utility. Figure~\ref{fig:reuters_mi} indicates that the MI adversary's advantage drops to the baseline level again for very weak DP guarantees of $\epsilon>10^2$ for all models. This behavior can be explained with the high amount of training examples and the smaller amount of hierarchical classes. Therefore, the gap between the empirically measured membership advantage and the upper bound on membership advantage diverge widely. 

For DBPedia in Figure~\ref{fig:dbpedia_utility}, the BoW model is again the most robust, and the Transformer model is least robust to the added noise during the training process, similar to the observations made on the BestBuy dataset. This is in line with our formulated expectations. The only exception are the measured utility metrics for $\epsilon\approx10^{-1}$, for which the BoW model performs worse than the CNN and Transformer model. MI metrics for the DBPedia HTC models are provided in Figure~\ref{fig:dbpedia_mi}. We see that the MI metrics for the BoW and Transformer models drop to the baseline level for very weak DP guarantees, similar to the Reuters models. Therefore, our MI adversary does by far not reach the theoretical upper $Adv$ bound. Notably, the MI metrics for the CNN model do not drop to the baseline level for the considered range for $z$ and resulting $\epsilon$. Hence, the gap between the measured $Adv$ and theoretical upper bound on $Adv$ reaches its lowest value for this model.

Overall, the privacy and utility results support our expectation that the utility of a high bias model such as BoW is less affected by the introduced noise than models with high variance such as Transformer. On the other hand, Transformer models are less prone to the MI attack due to better generalization which has generally been demonstrated aside from MI in related work~\cite{vaswaniAttentionAllYou2017,devlinBERTPretrainingDeep2018}.

\begin{table*}[ht!]
\captionsetup{justification=centering}
\centering
\begin{tabular}{cc|c|c|c|c|c|}
\cline{3-7}
                         &       & $n=41,625$,   & $n=41,625$, & $n=41,625$, & $n=4000$            & $n=400$           \\
                         &       & $14$ epochs  & $50$ epochs & $100$ epochs & $30$ epochs & $30$ epochs \\ \hline
\multicolumn{1}{|c|}{\multirow{4}{*}{$\cali{D}_{target}^{train}$}} & $L_1$ & $99.71\%$ & $99.94\%$ & $99.94\%$ & $99.94\%$ & $98.44\%$ \\ \cline{2-7} 
\multicolumn{1}{|c|}{}   & $L_2$ & $99.20\%$    & $99.86\%$         & $99.92\%$          & $99.44\%$            & $85.16\%$          \\ \cline{2-7} 
\multicolumn{1}{|c|}{}   & $L_3$ & $96.91\%$    & $99.74\%$         & $99.81\%$          & $96.07\%$            & $63.28\%$          \\ \cline{2-7} 
\multicolumn{1}{|c|}{}   & Loss  & $0.18$       & $0.01$            & $0.01$             & $0.30$               & $3.15$             \\ \hline
\multicolumn{1}{|c|}{\multirow{4}{*}{$\cali{D}_{target}^{test}$}}  & $L_1$ & $97.24\%$ & $96.93\%$ & $97.06\%$ & $93.47\%$ & $84.31\%$ \\ \cline{2-7} 
\multicolumn{1}{|c|}{}   & $L_2$ & $95.00\%$    & $94.79\%$         & $94.73\%$          & $87.89\%$            & $59.23\%$          \\ \cline{2-7} 
\multicolumn{1}{|c|}{}   & $L_3$ & $89.32\%$    & $91.11\%$         & $91.35\%$          & $78.53\%$            & $37.76\%$          \\ \cline{2-7} 
\multicolumn{1}{|c|}{}   & Loss  & $0.89$       & $1.60$            & $1.87$             & $2.08$               & $3.97$             \\ \hline
\multicolumn{2}{|c|}{$L_3$ Gap}  & $7.60\%$     & $8.63\%$          & $8.47\%$           & $17.54\%$            & $25.52\%$          \\ \hline
\multicolumn{2}{|c|}{Loss Ratio} & $5.2$        & $160$             & $187$              & $6.93$               & $1.26$             \\ \hline
\multicolumn{2}{|c|}{$Acc_{MI}$} & $53.06\%$    & $53.92\%$         & $54.01\%$          & $64.62\%$            & $75.00\%$          \\ \hline
\end{tabular}
\caption{Per-level accuracies and summarized loss for BestBuy  Transformer network without DP.}
\label{tab:target-model-mods}
\end{table*}

\subsection{Drivers for Attack Performance}
\label{sec:evaluation:effect:parameters}

Our experiments show that state-of-the-art MI attacks are not very effective when run against the trained HTC models. 
This leads to the question whether more HTC-specific attacks or less generalizing HTC target models would boost attack performance. In the following subsections, we first validate experimentally that the attack performance does not increase when introducing HTC-specific attack features. Second, we present several means that reduce target model generalization, where we show that especially reducing the number of training examples increases the vulnerability to MI attacks.

\subsubsection{HTC-specific attack model features}
\label{sec:evaluation:effect:attack}

We hypothesize that adapting the MI attack to exploit the hierarchical relation of the classes leads to an increased MI attack performance. Our approach to adapt the MI attack to HTC models is to extract additional attack features from the target model. We choose to evaluate two additional features in the following.

The former is a Boolean feature that we derive by checking if the HTC model's prediction before applying the post-processing step is consistent with the hierarchy of the HTC task. The latter is a scalar feature that is derived by multiplying the probabilities of the node with the highest softmax score on each level. The intuition behind this feature is to calculate a value that states how confident the target model is with the predicted label, since the assigned probability is obviously higher when the model outputs only small probabilities for the other labels. We refer to this attack feature as \textit{prediction confidence}.

We pass the features into the attack model with an additional FCN similar to the loss. We tested the effect of the additional features on all datasets and models. However, the results for this attack variant do not result in a significant change of the considered MI metrics (Figures~\ref{fig:bestbuy_mi_whf} to \ref{fig:dbpedia_mi_whf} in the appendix).

\subsubsection{Reduced Target Model Generalization}
\label{sec:evaluation:effect:target}

Next, we describe four approaches that we formulated to increase the attack model performance by reducing target model generalization in HTC. Each approach is first motivated and then evaluated based on experimental results.

\begin{figure}[ht!]
\captionsetup{justification=centering}
    \centering
    \begin{subfigure}{0.15\textwidth}
        \centering
        \includegraphics[width=\textwidth]{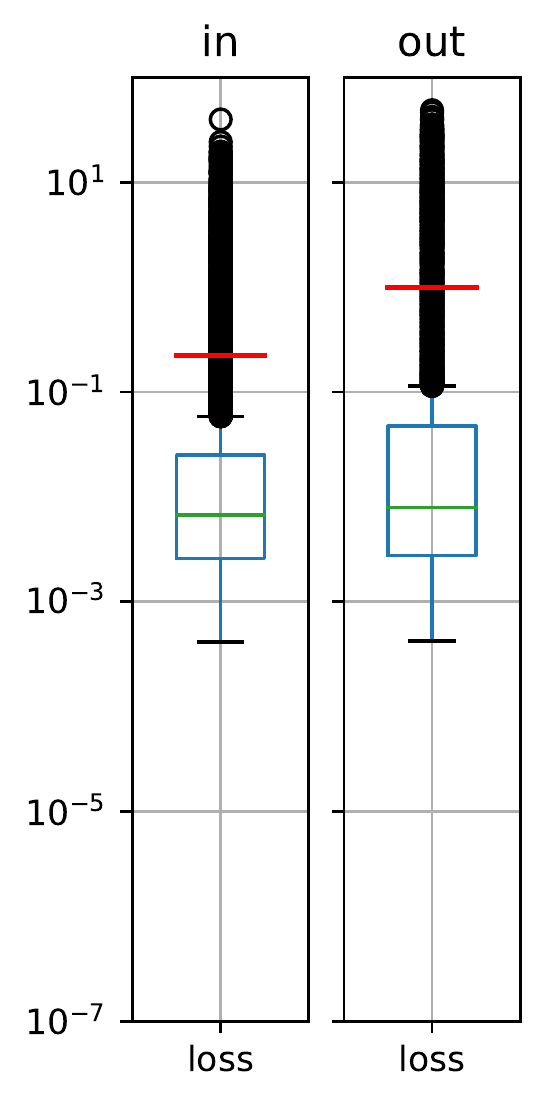}
        \caption{Loss on original model after 14 epochs}
        \label{fig:loss_dist:orig}
    \end{subfigure}%
    \begin{subfigure}{0.15\textwidth}
        \centering
        \includegraphics[width=\textwidth]{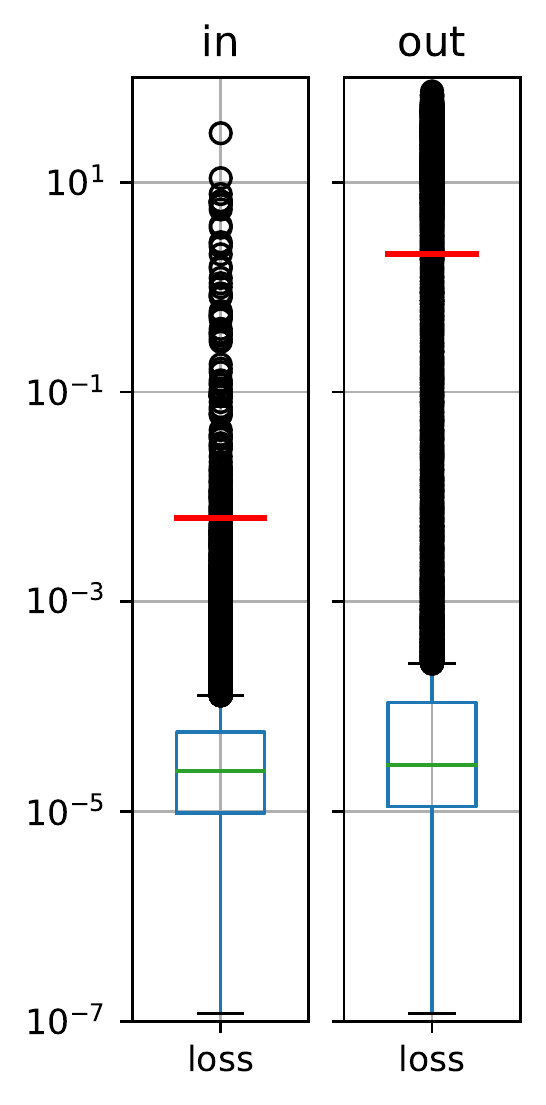}
        \caption{Loss on overfit model after 100 epochs}
        \label{fig:loss_dist:overfit}
    \end{subfigure}
    \caption{Loss distribution of members and non-members for BestBuy. Each boxplot is on a log scale depicting outliers (black), median (green) and mean (red) of the respective distribution.}
    \label{fig:loss_dist}
\end{figure}

First, we provoke an overfitted target model by training without early stopping for a fixed number of epochs, which is chosen significantly higher than the original number of epochs obtained with early stopping. In doing so, we deliberately force the model to overfit, i.e.,~adapt to the few samples in $\cali{D}_{target}^{train}$ instead of approximating the underlying distribution. We evaluated this approach based on the BestBuy Transformer classifier. Table~\ref{tab:target-model-mods} shows the metrics of the original model in the first column, which converged after 14 epochs. The second and third column reveal the metrics for overfit models, which are trained for 50 and 100 epochs, respectively. As expected, the overfit models achieve a smaller training loss and a higher test loss. However, surprisingly, the achieved test accuracy does not drop compared to the original model, while the training accuracy on $L_3$ increases to over $99\%$. The corresponding attack model accuracies rise from $53.06\%$ to $53.92\%$ and $54.01\%$ respectively. This insignificant change may appear counter-intuitive given the increased loss on $\cali{D}_{target}^{test}$. However, when analyzing the loss distribution of $\cali{D}_{target}^{train}$ and $\cali{D}_{target}^{test}$, we observe that the median losses decrease similarly as depicted in Figure~\ref{fig:loss_dist:orig} and \ref{fig:loss_dist:overfit}. The reason for the high average loss on $\cali{D}_{target}^{test}$ is due to the high loss value of a few outliers. Therefore, the loss ratio is not a consistently good indicator for MI attack effectiveness in practice and rather the accuracy gap should be taken into account.

\begin{figure*}[h]
\captionsetup{justification=centering}
    \centering
        \begin{subfigure}{0.3\textwidth}
        \centering
        \includegraphics[width=\textwidth]{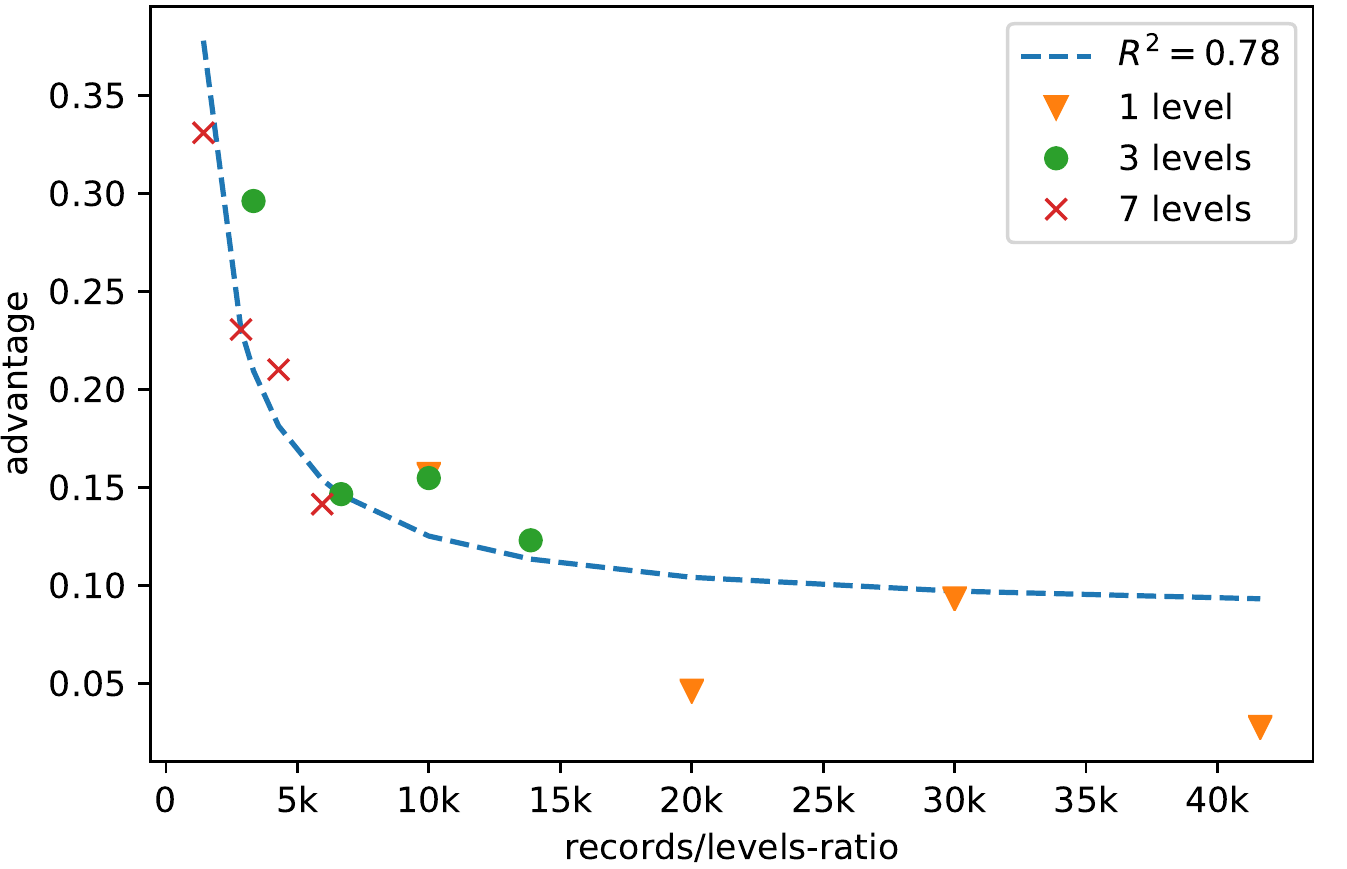}
        \caption{Relation between the records/levels ratio and attack advantage for BestBuy}
        \label{fig:bestbuy_cleaned_ratio}
    \end{subfigure}
    \begin{subfigure}{0.3\textwidth}
        \centering
        \includegraphics[width=\textwidth]{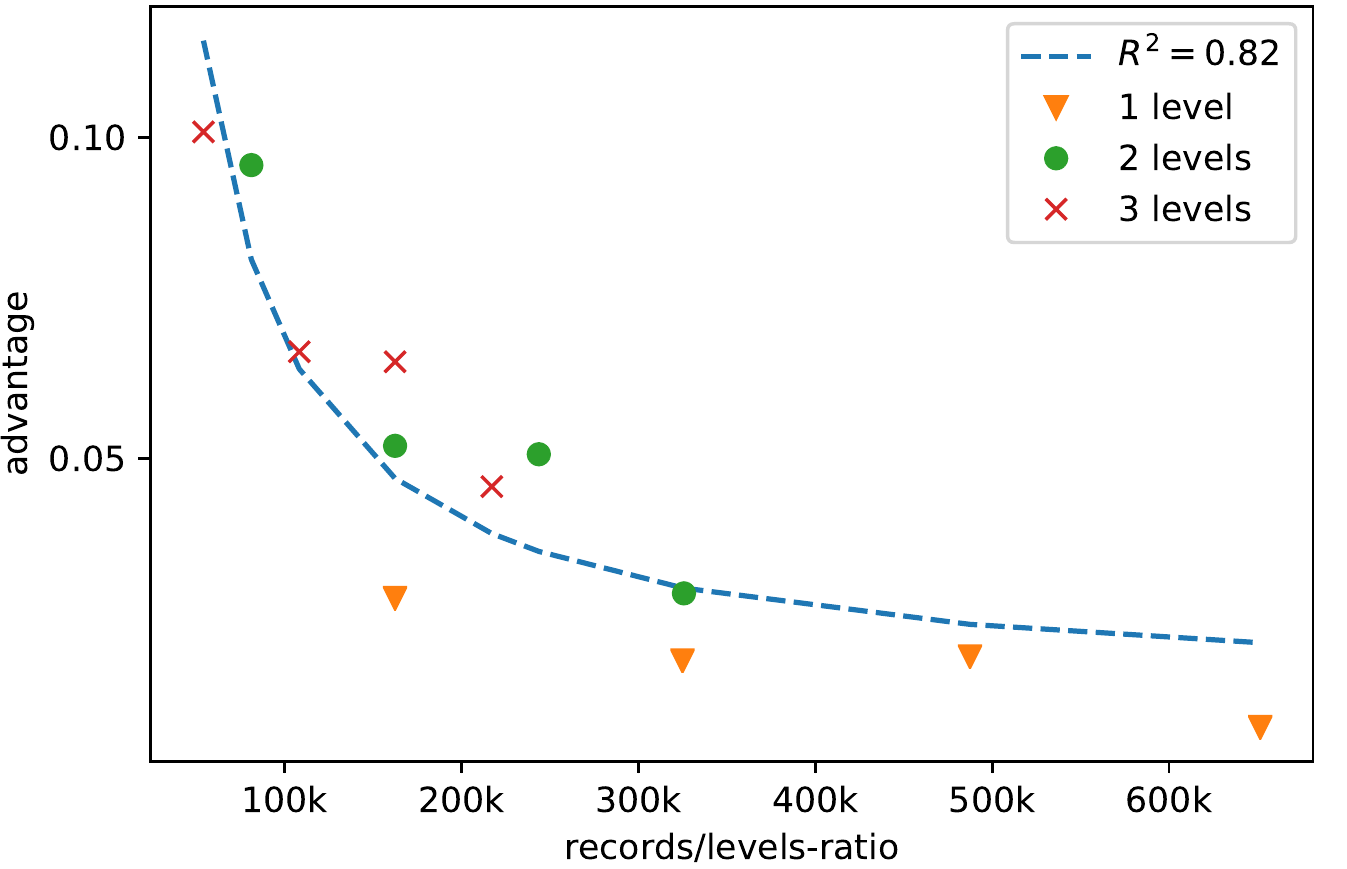}
        \caption{Relation between the records/levels ratio and attack advantage for Reuters}
        \label{fig:reuters_ratio}
    \end{subfigure}
    \begin{subfigure}{0.3\textwidth}
        \centering
        \includegraphics[width=\textwidth]{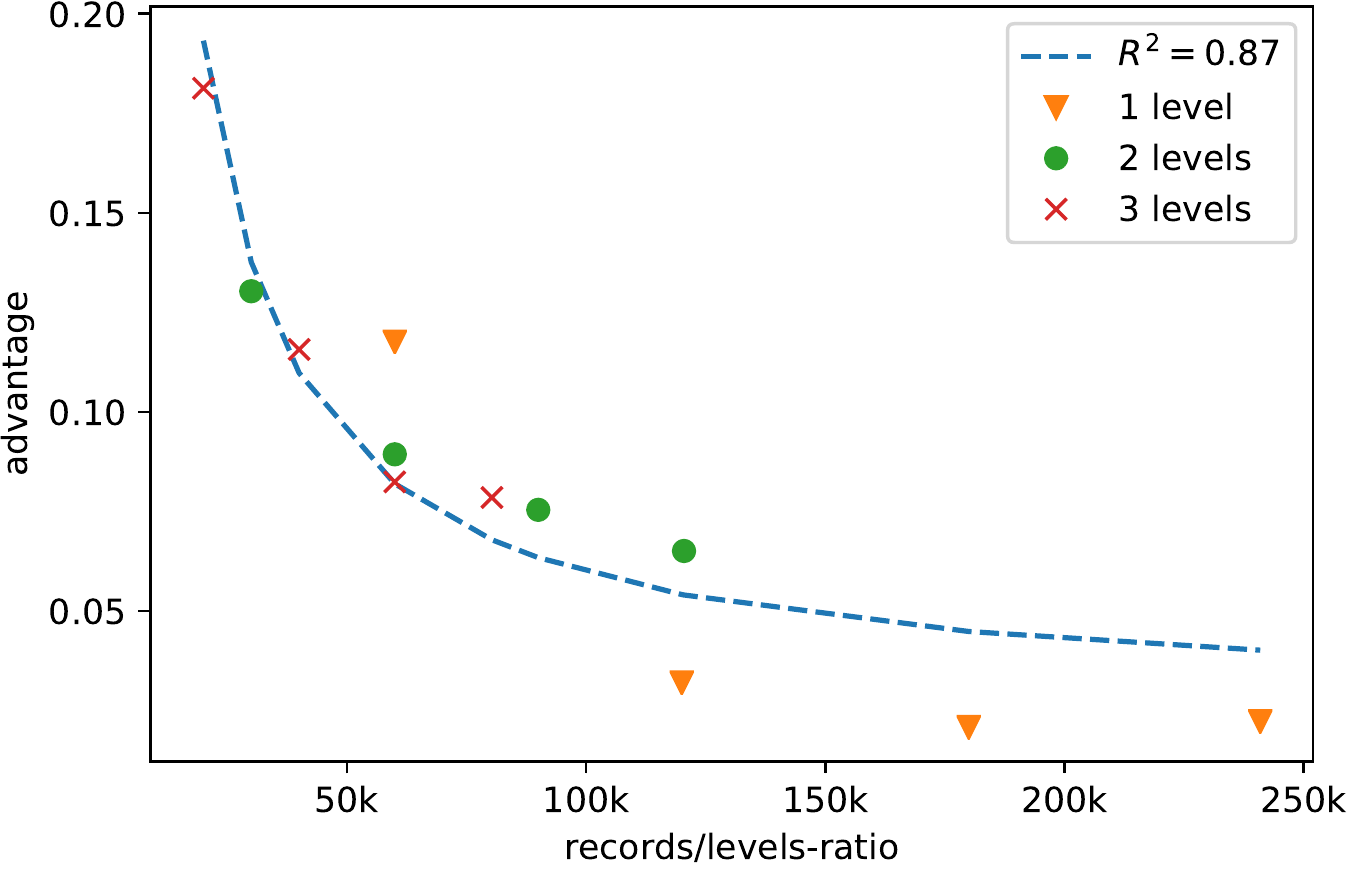}
        \caption{Relation between the records/levels ratio and attack advantage for DBPedia }
        \label{fig:dbpedia_ratio}
    \end{subfigure}
    \caption{Attack model advantage over ratio of records per hierarchy for an overall number of $\{0.25, 0.5, 0.75, 1\}\times n$ training records.}
    \label{fig:ratio}
\end{figure*}

Second, we reduce the number of training examples in $\cali{D}_{target}^{train}$. With this adaption, the hierarchical classifier should not generalize as well as the original classifier due to two reasons. First, the training dataset is less representative of the problem domain, and second, underrepresented classes contain even fewer examples. 
We again evaluate this approach based on the BestBuy transformer classifier, which originally contains $n=41,625$ training examples. Training the classifier with only $10\%$ of the training examples indeed leads to worse generalization with a maximum train-test gap of $17.54\%$ on the third level as shown in Table~\ref{tab:target-model-mods}. The trained attack model for this variation converged at $64.62\%$ accuracy, which is a significant increase compared to the original target model. Further reducing the training data to $n=400$ examples reduces the target model performance even more, with a maximum train-test gap of $25.52\%$ on the third level, as evident from Table~\ref{tab:target-model-mods}. For this target model with $n=400$, we observe an attack accuracy of $75\%$. In conclusion, reducing the number of training examples results in a significant MI attack improvement in comparison to changing the attack model features.

Third, we increase the number of hierarchy levels in the data, resulting in a more complex HTC task, which we again hypothesize to lead to worse generalization. Moreover, additional hierarchy levels lead to additional classification outputs and therefore additional attack features, which might further boost attack performance. And indeed, when increasing the number of levels from three to seven for the BestBuy BoW classifier, the corresponding MI attack accuracy rises from $56.15\%$ to $57.07\%$. To ensure that this effect was caused by noise, we also reduced the number of levels to one, resulting in an attack accuracy of $51.38\%$. This shows that an increased (decreased) number of hierarchy levels leads to higher (lower) attack performances.

Interestingly, when combining the effects of changing the number of hierarchy levels and decreasing the number of training examples, we made the observation that the ratio of the number of training examples to levels in the HTC task has an direct influence on the MI attack performance. We first noted this effect for BestBuy and then validated if we can also see it for the other datasets. The results confirm our expectations and are shown in Figure \ref{fig:ratio}.

Finally, we train the hierarchical classifier from scratch, without leveraging pre-trained weights to initialize the model. We hypothesize that this classifier variation might be more vulnerable to MI attacks, since a model without pre-training might tend to memorize more information about $\cali{D}_{target}^{train}$. Training the original BestBuy Transformer classifier from scratch did not converge to a useful HTC model, with only $18\%$ accuracy on the first level. This effect can be explained by the relatively small amount of training data compared to the large corpora the Transformer model is usually pre-trained on. The issue can be overcome by replacing the BERT-Base layers with BERT-Tiny layers, since tiny layer contain fewer weights to train. The hierarchical BERT-Tiny classifier trained from scratch yields a model with $75.17\%$ flat $Acc$. The trained attack model for this variation converged at $55.04\%$ accuracy, which is $\approx2\%$ higher than the attack on the original target model. This relatively small increase reveals that the use of pre-trained weights for the target models is not the reason for the relatively poor attack performance. 
\section{Discussion}
\label{sec:discussion}

\begin{figure*}[h]
    \centering
    \includegraphics[width=0.85\textwidth]{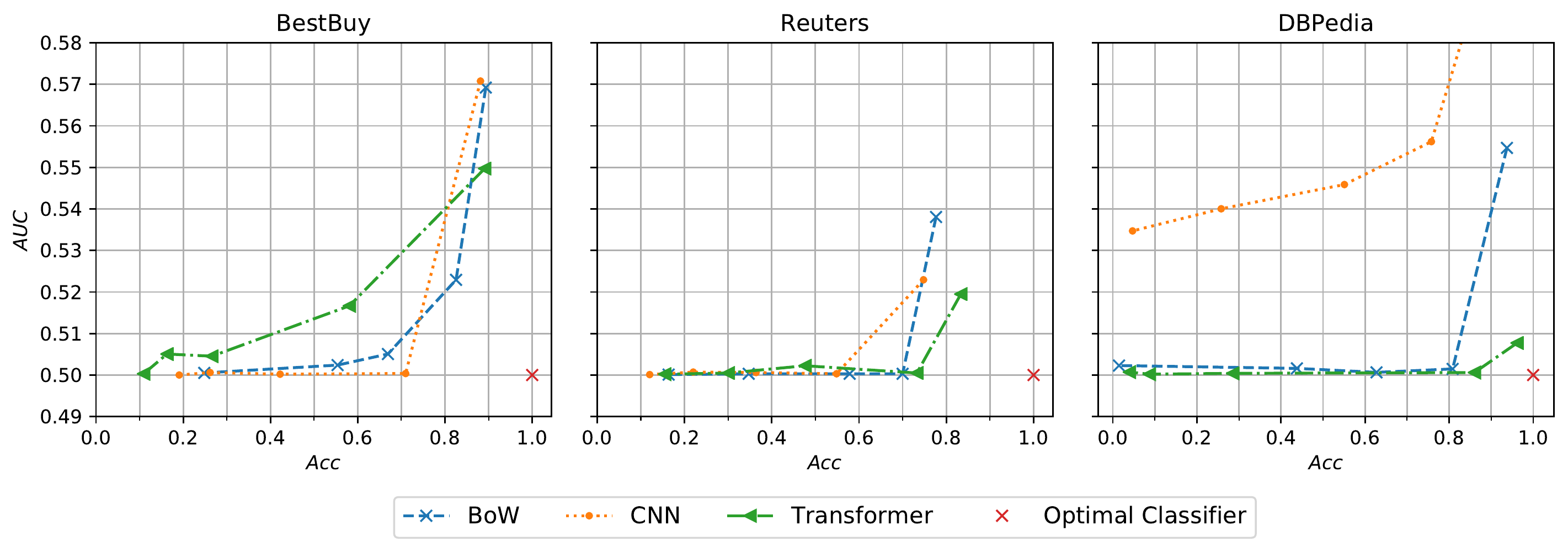}
    \caption[Privacy-utility trade-off per HTC model on each dataset.]{Privacy-utility trade-off per HTC model on each dataset. Privacy and utility are represented by MI $AUC$ and classification $Acc$, respectively. An optimal classifier would exhibit $100\%$ $Acc$ and no vulnerability to the MI adversary, expressed by $50\%$ $AUC$.}
    \label{fig:tradeoff}
\end{figure*}

\textbf{Large values for privacy parameter $\epsilon$ are sufficient to completely mitigate MI attacks with moderate decrease in model utility.} In our experiments, we enforced the HTC models to satisfy DP guarantees by clipping and perturbing the computed gradients during the training process. As expected, the experiments showed that enforcing DP in this way reduces the effectiveness of the performed MI attacks but also harms model utility. Figure~\ref{fig:tradeoff} summarizes the trade-off between classification accuracy and MI $AUC$ for each dataset. As can be seen, for all examined datasets, it is possible to completely mitigate the MI attack while reducing classification $Acc$ by $<20\%$. For BestBuy, the CNN model yields the best model utility for $AUC=50\%$. In contrast, for Reuters and DBPedia, the Transformer yields the best model utility for $AUC=50\%$. This may be explained by the size and average text length of the datasets, since for small datasets with short texts (e.g., BestBuy) the CNN model is well suited, while for larger datasets with longer texts (e.g., Reuters, DBPedia) the Transformer architecture is suited better.

\textbf{Overfitting is a key driver for MI attack performance.}
Our experiments support the understanding that overfitting leads to higher MI attack performance. We therefore extend the findings of previous work for non-textual datasets to HTC~\cite{shokriMembershipInferenceAttacks2017, nasrComprehensivePrivacyAnalysis2018}.
To prevent overfitting, our analysis of drivers for MI attack performance suggests to gather as many training examples as possible and only predict as many hierarchy levels as needed. Adding HTC-specific features to the attack model did not increase MI attack performance, confirming the validity of our results also for stronger adversaries.

\textbf{Similar DP privacy parameters do not imply a similar MI attack effectiveness.} The experimental results show that the empirical MI risk for similar DP guarantees varies within each dataset but also within each model architecture. Therefore, we can again summarize that the MI attack effectiveness depends on the chosen model architecture and the dataset. Unfortunately, the results do not point to a model architecture that is strictly better suited to mitigate the MI attack. However, we recommend using a model with relatively few parameters such as the BoW model for smaller datasets, whereas for larger datasets models with a high number of parameters such as the Transformer classifier yields a favorable trade-off.

\textbf{The BoW model's utility was reduced least by the added DP noise.} Across all datasets, we observed that the CNN and Transformer model's utility scores were impacted more heavily compared to the BoW model's utility for similar DP guarantees. On two of the three datasets, the Transformer model's utility is impacted even more severely than the CNN model's utility. This finding suggests that a higher number of weights in an ANN might correlate with a stronger impact of DP training on the ANN utility. Specifically, the number of ANN weights is lowest in the BoW model and highest in the Transformer model. This insight should be taken into account when a data scientist wants to train an ANN based on a given formal DP guarantee.

\textbf{HTC ANNs exhibit a big gap between empirical and theoretical MI risk}. The obtained results support the conclusions by Jayaraman et al.~\cite{jayaramanEvaluatingDifferentiallyPrivate2019}, who found that there remains a big gap between what state-of-the-art MI attacks can infer and what is the maximum that can theoretically be inferred according to the bound presented by Yeom et al.~\cite{yeomPrivacyRiskMachine2018}. During evaluation, we measured the membership advantage and compared it to the theoretical membership advantage bound, which can be calculated given the respective DP guarantee. We showed that this conclusion also holds in the context of HTC. 
\section{Related work}
\label{sec:related_work}

This work is related to HTC, DP in NLP and MI attacks for evaluating the privacy of DP ML models. Therefore, in this section, we briefly introduce the most relevant publications in the respective research fields.

Stein et al.~\cite{steinAnalysisHierarchicalText2019} analyze the performance of different hierarchical text classifiers on the Reuters (RCV1) dataset that we also use for the experiments in our work. The authors find that a fastText-based classifier works better than a CNN-based classifier initialized with the same embeddings. For evaluation, all possible types of metrics are used in the paper, namely flat, hierarchical, and LCA metrics. 
Interestingly, the authors do not consider any Transformer-based HTC model, even though Transformer architectures represent state-of-the-art for text classification.

Abadi et al.~\cite{abadiDeepLearningDifferential2016} formulate an implementation of the DP stochastic gradient descent, which uses the Gaussian mechanism to perturb gradient descent optimizers for ANNs. As ANNs are widely used in modern NLP and natural language data in many cases contain sensitive data, there are various publications regarding DP in NLP. While Vu et al.~\cite{vuDpUGCLearnDifferentially2019} learn DP word embeddings for user generated content, so that the resulting word embeddings can be shared safely, we focus on safe sharing of whole ML models. Other works use DP for author obfuscation in text classification~\cite{fernandesGeneralisedDifferentialPrivacy2019, weggenmannSynTFSyntheticDifferentially2018}. In contrast, our work addresses the privacy-utility trade-off for perturbation of the gradient descent optimizer.

Carlini et al.~\cite{carliniSecretSharerEvaluating2019} successfully apply DP to prevent information leakage in a generative model, specifically an ANN generating text. They introduce the \textit{exposure} metric to measure the risk of unintentionally memorizing rare or unique training-data sequences in generative models. Our work does not consider generative models, but solely classification models.

Empirical MI attacks against machine learning models such as the attack used in this work were first formulated by Shokri et al.~\cite{shokriMembershipInferenceAttacks2017} in the form of black-box MI. The authors compare MI attacks with model inversion attacks, which abuse access to an ML model to infer certain features of the training data. In contrast to model inversion attacks, MI attacks target a specific training example instead of targeting all training examples for a specific class. Therefore, the authors argue that successful MI attacks indicate unintended information leakage. Misra~\cite{misraBLACKBOXATTACKS2019a} uses black-box MI attacks to assess the information leakage of generative models.

Nasr et al.~\cite{nasrComprehensivePrivacyAnalysis2018} showed that white-box MI attacks, that take the target model's internal parameters into account, are more effective than black-box MI attacks. Additionally, the authors assume that the adversary owns a fraction of the data owner's sensitive data. This stronger assumption about the adversary's knowledge increases the overall strength of the MI attack compared with black-box MI attacks.

While Rahman et al.~\cite{rahmanMembershipInferenceAttack2018} analyze the effect of different values for $\epsilon$ on the effectiveness of only black-box MI attacks, Bernau et al.~\cite{bernauAssessingDifferentiallyPrivate2020} take both black-box and white-box attacks into account. However, both publications mostly consider specifically crafted non-textual MI datasets. We consider real-world textual training data.

Yeom et al.~\cite{yeomPrivacyRiskMachine2018} introduce membership advantage to measure the success of an MI attack. Furthermore, they formulate a theoretical upper bound for the membership advantage that depends on the DP guarantees of the target model. Humphries et al.~\cite{Humphries} derive a bound for membership advantage that is tighter than the bound derived by Yeom et al. by analyzing the impact of giving up the i.i.d.~assumption. However, there is a gap between the theoretic upper bound for the membership advantage and the membership advantage of state-of-the-art MI attacks, as has been shown by Jayaraman et al.~on numeric and image data~\cite{jayaramanEvaluatingDifferentiallyPrivate2019}. In our work, we investigate this gap in the context of ANNs for HTC. 
\section{Conclusion}
\label{sec:conclusion}

This work analyzed and compared the privacy-utility trade-off in DP HTC under a white-box MI adversary. Even without the use of DP, white-box MI attacks only posed a minor risk to the three HTC model architectures and reference dataset that we considered. In consequence, large privacy parameters were sufficient to fully mitigate the white-box attack. We particularly observed Transformer-based HTC models to be rather resistant to MI attacks without using DP at all. 

However, the privacy-accuracy trade-off for full mitigation of the white-box MI attack is differing widely for all considered models and datasets. Our results suggest that the Transformer model is also favorable for large datasets with long texts when using DP, while the CNN model is favorable for smaller datasets with shorter texts. However, if hardware costs shall be minimized or the training examples shall be protected with a strong formal DP guarantee (i.e., small $\eps$ value), the fastText based BoW model is a good choice due to the high robustness against DP perturbation. Our experiments also confirm a large gap between the empirical membership advantage of the MI white-box attack and the theoretical DP membership advantage bound for HTC datasets and models.
\section{Acknowledgements}
\label{sec:ack}
We would also like to thank the anonymous reviewers for their immensely helpful suggestions to improve the readability and contents of this paper.
This work has received funding from the European Union’s Horizon 2020 research and innovation program under grant agreement No. 825333 (MOSAICROWN).
The project that gave rise to these results received the support of a fellowship from ”la Caixa” Foundation (ID 100010434) and from the European Union’s Horizon 2020 research and innovation programme under the Marie Skłodowska-Curie grant agreement No 847648. The fellowship code is LCF/BQ/PR20/11770009.
The work of Javier Parra-Arnau has been supported through an Alexander von Humboldt Post-Doctoral Fellowship.

\bibliographystyle{abbrv}
\bibliography{DPHTC}

\onecolumn
\appendix
\section*{Appendix: Additional Figures}
\label{app:figures}
\begin{figure*}[ht!]
\captionsetup{justification=centering}
    \centering
    \begin{subfigure}{0.75\textwidth}
        \centering
        \includegraphics[width=\textwidth]{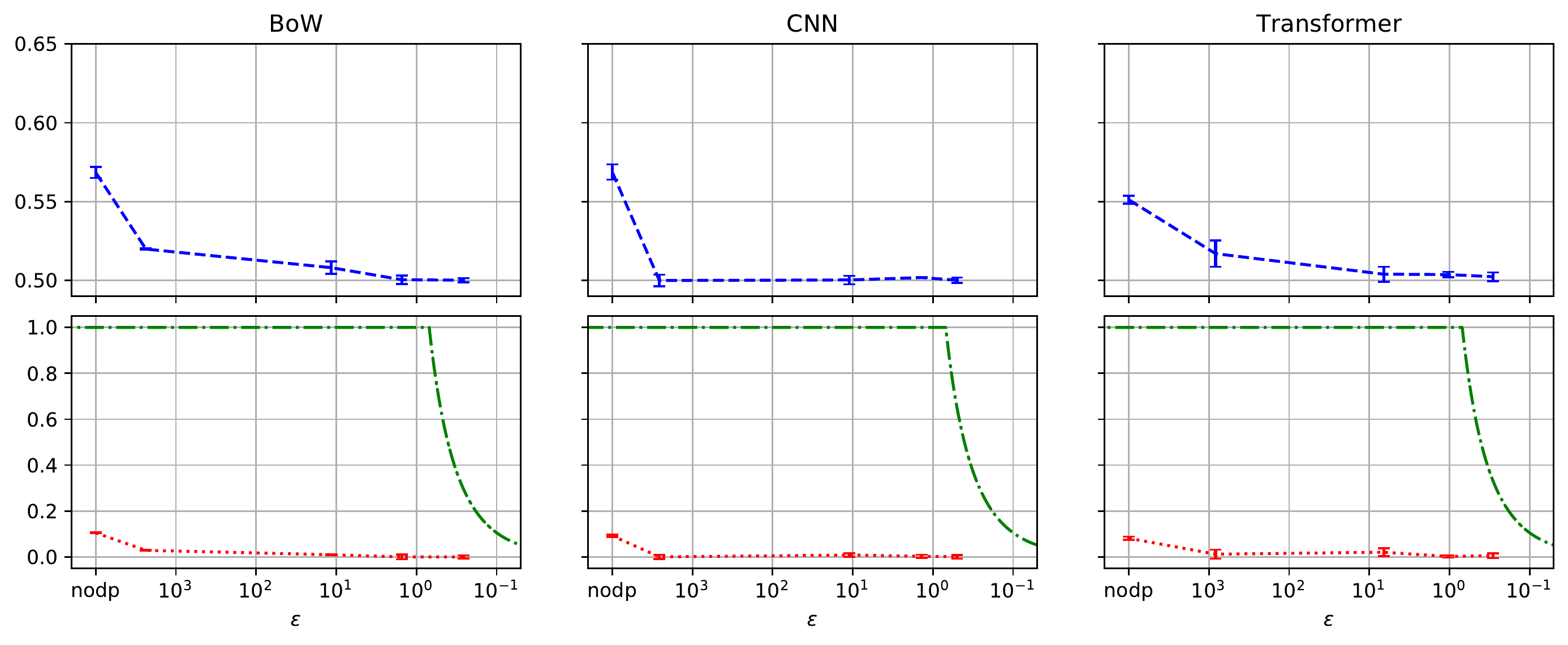}
        \caption{MI against BestBuy over $\eps$}
        \label{fig:bestbuy_mi_whf}
    \end{subfigure}
    \begin{subfigure}{0.75\textwidth}
        \centering
        \includegraphics[width=\textwidth]{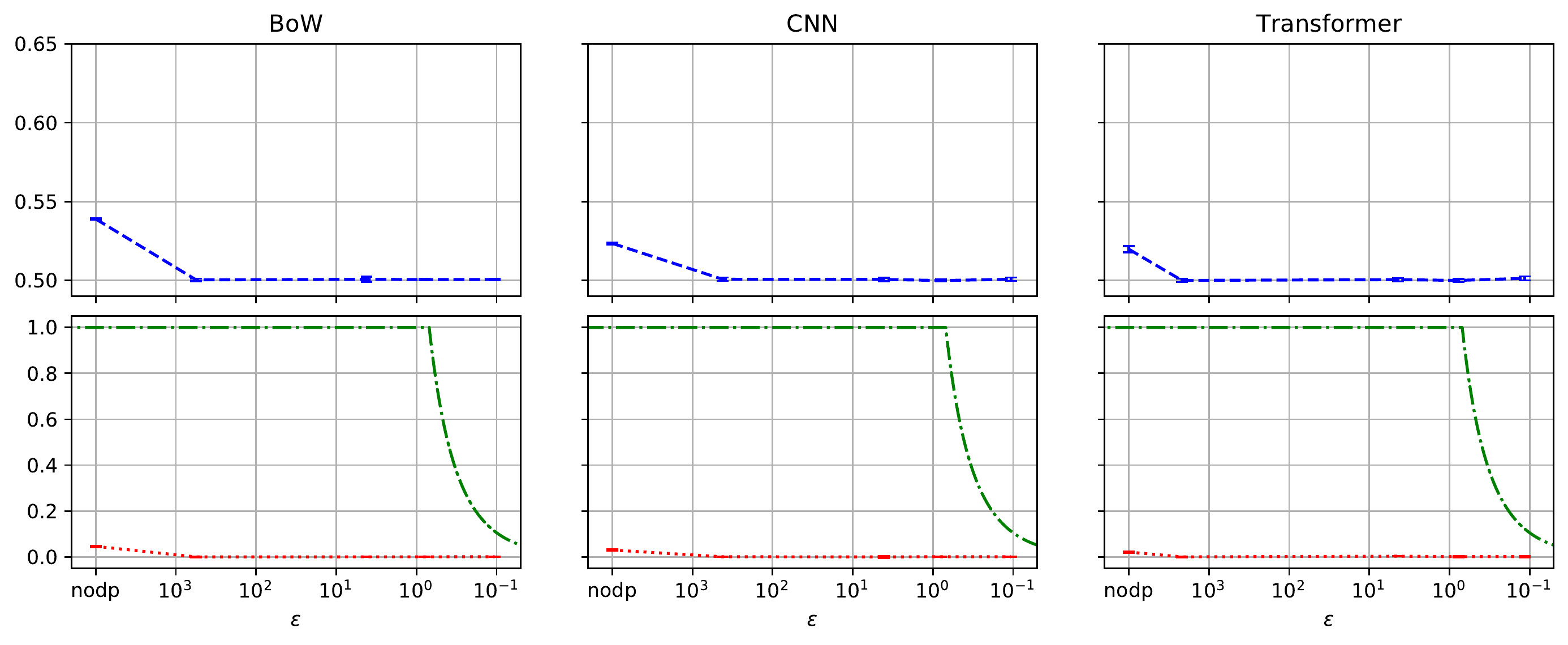}
        \caption{MI against Reuters over $\eps$}
        \label{fig:reuters_mi_whf}
    \end{subfigure}
    \begin{subfigure}{0.75\textwidth}
        \centering
        \includegraphics[width=\textwidth]{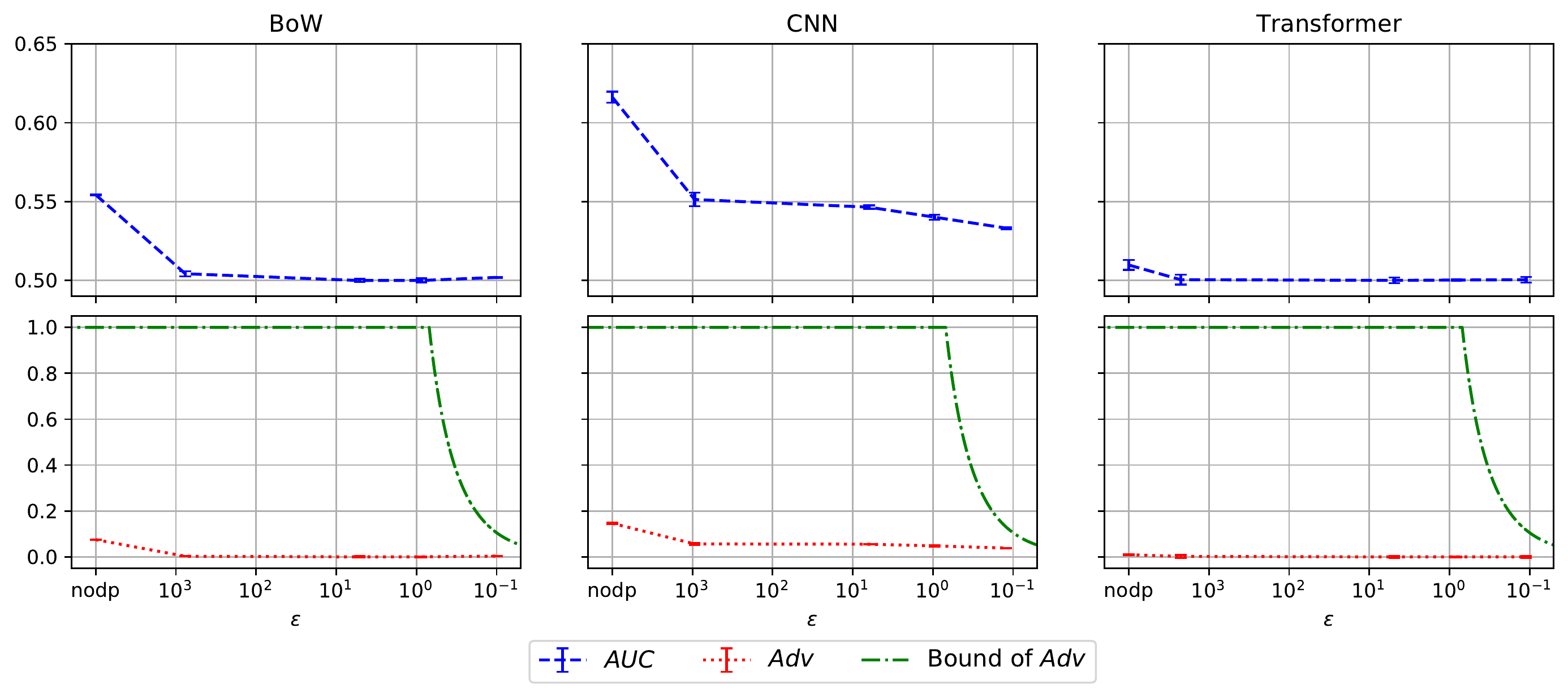}
        \caption{MI against DBPedia over $\eps$}
        \label{fig:dbpedia_mi_whf}
    \end{subfigure}
    \caption{MI AUC, $Adv$ and Bound on MI $Adv$ per dataset with additional hierarchical attack features}
    \label{fig:appendix}
\end{figure*}

\end{document}